\documentclass[ALICE,manyauthors]{cernphprep}

\newcommand{\s}{\ensuremath{\sqrt{s}}\xspace}
\newcommand{\GeVc}{\ensuremath{\textrm{GeV}/c}\xspace}
\newcommand{\GeVcc}{\ensuremath{\textrm{GeV}/c^2}\xspace}
\newcommand{\pt}{\ensuremath{p_{\rm T}}\xspace}
\newcommand{\mt}{\ensuremath{m_{\rm T}}\xspace}
\newcommand{\kt}{\ensuremath{k_{\rm T}}\xspace}

\newcommand{\ptrange}[2]{$#1<\pt<#2~\textrm{GeV}/c$}
\newcommand{\dEdx}{\ensuremath{\textrm{d}E/\textrm{d}x}\xspace}
\newcommand{\sNN}{\ensuremath{\sqrt{s_{\rm NN}}~}}
\newcommand{\RpPb}{\ensuremath{R_{\rm pPb}}\xspace}
\newcommand{\RPbPb}{\ensuremath{R_{\rm PbPb}}\xspace}
\newcommand{\RpA}{\ensuremath{R_{\rm pA}}\xspace}
\newcommand{\RAA}{\ensuremath{R_{\rm AA}}\xspace}
\newcommand{\RdAu}{\ensuremath{R_{\rm dAu}}\xspace}

\newcommand{\pPb}{\mbox{p--Pb~}}
\newcommand{\pp}{\mbox{pp~}}
\newcommand{\PbPb}{\mbox{Pb--Pb}\xspace}
\newcommand{\AuAu}{\mbox{Au--Au}\xspace}
\newcommand{\AAc}{AA\xspace}
\newcommand{\pA}{p--A\xspace}

\newcommand{\nsigmaTPC}{\ensuremath{n_{\sigma}^{\textrm{TPC}}}\xspace}
\newcommand{\nsigmaTOF}{\ensuremath{n_{\sigma}^{\textrm{TOF}}}\xspace}

\newcommand{\sqpA}{\ensuremath{\sqrt{\textit{s}_{\rm NN}}} = 5.02~TeV\xspace}
\newcommand{\sqAA}{\ensuremath{\sqrt{\textit{s}_{\rm NN}}} = 2.76~TeV\xspace}

\newcommand{\smbeq}{\sigma_{\rm mb}^{\rm V0}}
\newcommand{\jpsi}{J/$\psi$\xspace}
\newcommand{\KL}{$\rm{K}_{\rm L}^{0}$\xspace}
\newcommand{\KS}{$\rm{K}_{\rm S}^{0}$\xspace}
\newcommand{\dminabs}{$|d_{0}|$\xspace}
\newcommand{\dmin}{$d_{0}$\xspace}
\usepackage[comma,square,numbers,sort&compress]{natbib}
\usepackage{lineno}

\usepackage{hyperref}
\usepackage{graphicx}  
\usepackage{dcolumn}   
\usepackage{bm}        
\usepackage{amssymb}   
\usepackage{amsfonts}
\usepackage{graphics}
\usepackage{grffile}   
\usepackage{epsfig}
\usepackage{units}
\usepackage[usenames]{color}
\usepackage[normalem]{ulem} 
\usepackage[utf8]{inputenc}
\usepackage[T1]{fontenc}
\usepackage{subfigure}

\begin{document}

\begin{titlepage}
\PHyear{2016}
\PHnumber{222}      
\PHdate{13 September}  
%

\title{Measurement of electrons from beauty-hadron decays in \\
 \pPb collisions at $\mathbf{\sqpA}$ and \\ \PbPb collisions at $\mathbf{\sqAA}$}
\ShortTitle{Measurement of electrons from beauty-hadron decays in \pPb and \PbPb}   

\Collaboration{ALICE Collaboration%
         \thanks{See Appendix~\ref{app:collab} for the list of collaboration members}}
\ShortAuthor{ALICE Collaboration}       
\begin{abstract}

The production of beauty hadrons was measured via semi-leptonic decays at mid-rapidity with the ALICE detector at the LHC in the transverse momentum interval \ptrange{1}{8} in minimum-bias \pPb collisions at \sqpA and in \ptrange{1.3}{8} in the 20\% most central \PbPb collisions at \sqAA. The pp reference spectra at \s~=~5.02~TeV and \s~=~2.76~TeV, needed for the calculation of the nuclear modification factors \RpPb and \RPbPb, were obtained by a pQCD-driven scaling of the cross section of electrons from beauty-hadron decays measured at \s~=~7~TeV.
In the \pt interval \ptrange{3}{8} a suppression of the yield of electrons from beauty-hadron decays is observed in \PbPb compared to pp collisions. Towards lower \pt, the \RPbPb values increase with large systematic uncertainties. The \RpPb is consistent with unity within systematic uncertainties and is well described by theoretical calculations that include cold nuclear matter effects in \pPb collisions. The measured \RpPb and these calculations indicate that cold nuclear matter effects are small at high transverse momentum also in \PbPb collisions. Therefore, the observed reduction of \RPbPb below unity at high \pt may be ascribed to an effect of the hot and dense medium formed in \PbPb collisions.

\end{abstract}
\end{titlepage}
\setcounter{page}{2}

\section{Introduction}\label{Intro}

In collisions of heavy nuclei at ultra-relativistic energies, a high-density colour-deconfined state of strong\-ly-interacting matter, called Quark--Gluon Plasma (QGP), is expected to be produced~\cite{Borsanyi:2010bp,Bazavov:2014pvz}. 
Due to their large masses ($m_{\rm Q} \gg \Lambda_{\rm QCD}$), heavy quarks (charm and beauty) are almost exclusively produced in the early stage of the collision via hard parton scatterings characterised by production-time scales of less than 0.1 and 0.01~fm/$c$ for charm and beauty quarks, respectively~\cite{Anton:2015}. They can, therefore, serve as probes to test the mechanisms of medium-induced parton energy loss, because the formation time of the QGP medium is  expected to be about 0.3 fm/$c$~\cite{Liu:2012ax} and its decoupling time is about 10~fm/$c$ for collisions at LHC energies~\cite{Aamodt:2011mr}. 
Due to their stronger colour coupling to the medium gluons are argued to lose more energy than quarks~\cite{Gyulassy:1993hr,Baier:1996sk,Wiedemann:2000za}. Furthermore, the radiative energy loss of heavy quarks is predicted to be reduced with respect to light quarks due to the mass-dependent restriction of the phase space into which medium-induced gluon radiation can take place (dead-cone effect)~\cite{Dokshitzer:2001zm,Armesto:2003jh,PhysRevLett.93.072301,Djordjevic:2003zk}. 
The effect of the charm-quark mass on energy loss becomes negligible at high transverse momentum, \pt $\gtrsim$ 10~\GeVc, where the ratio \mbox{${m_{\rm c}}/$\pt} approaches zero~\cite{Armesto:2005iq}. Therefore, due to the larger mass, beauty quarks can be sensitive probes for testing the mass dependence of the parton energy loss up to transverse momenta well above 10~\GeVc~\cite{Armesto:2005iq}. Final-state effects, such as colour-charge and mass dependence of parton energy loss, can be studied experimentally through the spectra of hadrons containing heavy quarks in comparison with light-flavour hadrons in heavy-ion (\AAc) collisions. 

The understanding of final-state effects requires measurements of initial-state effects in Cold Nuclear Matter (CNM), which are inherent to nuclei in the collision system and thus present in \AAc collisions. Measurements in proton--nucleus (\pA) collisions are used to investigate cold nuclear matter effects such as the modification of the Parton Distribution Functions (PDF) inside the nucleus with respect to those in the proton, \kt broadening via parton collisions inside the nucleus prior to the hard scattering and energy loss in cold nuclear matter~\cite{Armesto:2006ph,Fujii:2006ab,Cronin:1974zm,Sharma:2009hn,Kang:2014hha}. 
The effects of hot (cold) nuclear matter can be studied using the nuclear modification factor, \RAA (\RpA), defined as the ratio of the \pt distributions measured in \AAc (\pA) collisions with respect to the one in pp collisions:
\begin{equation}
\RAA = \frac{1}{\langle T_{AA} \rangle}\frac{{\rm d}N_{\rm AA}/{\rm d}\pt}{{\rm d}\sigma_{\rm pp}/{\rm d}\pt}   ,
\label{RAAFormula}
\end{equation}
where ${\rm d}N_{\rm AA}/{\rm d}\pt$ and ${\rm d}\sigma_{\rm pp}/{\rm d}\pt$ are the \pt-differential yield and production cross section of a given particle species in \AAc and pp collisions, respectively, and $\langle T_{\rm AA} \rangle$ is the average of the nuclear overlap function for the centrality range under study~\cite{Miller:2007ri}.  

Previous beauty-hadron production measurements in pp collisions at various energies at RHIC~\cite{Aggarwal:2010xp,Agakishiev:2011mr}, the Tevatron~\cite{PhysRevD.71.032001} and the LHC~\cite{Anton:2015,Abelev:2012sca,Abelev:2014gla,Abelev:2012gx,Aaij:2012jd,Chatrchyan:2011vh,Aad:2011sp} are described by Fixed Order plus Next-to-Leading-Log perturbative Quantum Chromodynamics (FONLL pQCD) calculations ~\cite{Cacciari:1998it,Cacciari:2001td,Cacciari:2012ny} within uncertainties.

At both RHIC and the LHC, a suppression of the yield of D mesons and high-\pt electrons and muons from heavy-flavour hadron decays was observed in \AAc collisions. The suppression is nearly as large as that of light-flavour hadrons at high \pt~\cite{Adare:2006nq,Adare:2010de,Adam:2015nna,Adam:2015sza,Abelev:2012qh}. The D meson and pion \RPbPb were found to be consistent within uncertainties and described by model calculations that include a colour-charge dependent energy loss~\cite{Adam:2015nna,Abelev:2014laa,Djordjevic:2014tka}. However, in addition to energy loss, the nuclear modification factor is also influenced by e.g.\ the parton \pt spectrum and the fragmentation into hadrons~\cite{Djordjevic:2013pba,Armesto:2005iq}.
Furthermore, the nuclear modification factors \RPbPb of prompt D mesons and of \jpsi from B meson decays were compared in the \pt interval \ptrange{8}{16} for \mbox{D mesons} and \ptrange{6.5}{30} for \jpsi mesons in order to have a similar average \pt ($\approx$10 \GeVc) for the heavy hadrons~\cite{Adam:2015nna,Adam:2015rba,Chatrchyan:2012np}. This comparison with models indicates that charm quarks lose more energy than beauty quarks in this \pt interval in central \PbPb collisions. The b-jet yield as measured in \PbPb collisions also shows a suppression compared with the yield expected from pp collisions in the jet-\pt interval \ptrange{70}{250} \cite{Chatrchyan:2013exa}. 
Recently, the relative contributions of electrons from charm- and beauty-hadron decays were measured as a function of transverse momentum in \AuAu collisions at RHIC~\cite{Adare:2015hla}. There is a hint that in the momentum interval \ptrange{3}{4} the $R_{\rm AuAu}$ of electrons from beauty-hadron decays is larger than that of electrons from charm-hadron decays.

In \pPb collisions at the LHC, the nuclear modification factors of B mesons~\cite{Khachatryan:2015uja}, b-jets~\cite{CMS:2014tca}, \jpsi from beauty-hadron decays~\cite{CMS:2015qud,Aaij:2013zxa}, leptons from heavy-flavour hadron decays and D mesons ~\cite{Abelev:2014hha,Adam:2015qda} were investigated extensively. The results are consistent with unity within uncertainties and compatible with theoretical calculations including cold nuclear matter effects~\cite{CMS:2014tca,Aaij:2013zxa,CMS:2015qud,Abelev:2014hha}. Therefore, the observed suppression of charm and beauty yields at high \pt in \PbPb collisions is not explained in terms of initial-state effects but is due to strong final-state effects induced by hot partonic matter.  

In central d--Au collisions at \sNN~=~200~GeV at RHIC, an enhancement was measured at backward rapidity by means of \RdAu of muons from heavy-flavour hadron decays~\cite{Adare:2013lkk}. Theoretical calculations including modified PDFs cannot describe the data, implying that models incorporating only initial-state effects are not sufficient and suggesting the possible importance of final-state effects in the d--Au collision system. Recently, a potential signature of collective behaviour in small systems was observed via the anisotropic flow parameter ${v_{\rm 2}}$ of charged hadrons in \pPb collisions~\cite{Abelev:2012ola,ABELEV:2013wsa,Aad:2013fja,CMS:2012qk} and in d--Au collisions\cite{Adare:2013piz,Adamczyk:2015xjc}, suggesting radial flow as a possible explanation of the enhancement of the \RdAu~\cite{Sickles:2013yna}.

In this paper, the invariant cross section in \pPb and yield in \PbPb collisions are presented together with the nuclear modification factors, \RpPb and \RPbPb, of electrons from beauty-hadron decays in \pPb and \PbPb collisions at \sqpA and \sqAA, respectively. The identification of electrons from beauty-hadron decays is based on their separation from the interaction vertex, induced by the sizable lifetime of beauty hadrons. The \pPb (\PbPb) measurement covers the rapidity range $|y_{\rm lab}| \le 0.6 $ ($|y_{\rm lab}| \le 0.8 $) and the \pt interval \ptrange{1.0}{8.0} (\ptrange{1.3}{8.0}). In the \pPb collisions, due to the different energy per nucleon of the proton and lead beam, the centre-of-mass system (cms) is shifted by $\Delta y = 0.465 $ in the proton beam direction, resulting in the rapidity coverage $-1.06 < y_{\rm cms} < 0.14$ for electrons. Given the cms energies and the rapidity coverages in the \pPb and \PbPb collisions, both measurements probe, at the lowest \pt, similar values of Bjorken-$x$ of about $10^{-3}$ for electrons from beauty-hadron decays~\cite{Dainese:2003zu}. The \PbPb measurement is restricted to the 20\% most central \PbPb collisions, where the largest effect of energy loss on heavy-flavour production is expected. 

The paper is organised as follows: Section~\ref{Exp} describes the experimental apparatus and the data samples used in both analyses, which are outlined in Section~\ref{Ana}. Details of the analysis in \pPb and \PbPb collisions are given in Sections~\ref{AnapPb} and~\ref{anaPbPb}, respectively. The determination of the pp reference spectra for the calculations of the \RpPb and \RPbPb is reported in Section~\ref{ppref}. The results are presented and discussed in Section~\ref{Res}. Section~\ref{Summarytex} summarises the results.

\section{Experimental apparatus and data samples}\label{Exp}

A comprehensive description of the ALICE apparatus and its performance can be found in~\cite{Aamodt:2008zz} and~\cite{Abelev:2014ffa}, respectively. Electron tracks were reconstructed and identified using detectors located inside the solenoid magnet that generates a field of 0.5~T parallel to the beam direction. Forward and backward detectors inside and outside the magnet were employed for triggering, background rejection and event characterisation.

Charged particles are tracked with the Inner Tracking System (ITS)~\cite{Aamodt:2008zz,Aamodt:2010aa} and the Time Projection Chamber (TPC)~\cite{Alme:2010ke} in the pseudorapidity range $|\eta|<$ 0.9. The ITS consists of six cylindrical layers of silicon detectors. The two innermost layers are made of Silicon Pixel Detectors (SPD), the two middle layers of Silicon Drift Detectors (SDD) and the two outermost layers of Silicon Strip Detectors (SSD). In the direction perpendicular to the detector surface, the total material budget of the ITS corresponds on average to 7.7\% of a radiation length~\cite{Aamodt:2010aa}. In this analysis, the ITS was also used to reconstruct the primary (interaction) vertex and the track impact parameter \dmin, defined as the distance of closest approach of the track to the interaction vertex in the plane transverse to the beam direction. The resolution on \dmin is better than 65~$\mu$m and 70~$\mu$m for charged particles with momenta larger than 1~\GeVc in \PbPb and \pPb collisions~\cite{Abelev:2014ffa}, respectively, including the resolution of the primary vertex determination. The particle identification capability of the four outer layers of the ITS via the measurement of the ionisation energy loss \dEdx was used at low transverse momentum in the \pPb analysis. The TPC, which provides up to 159 space points per track, is used for particle identification via the measurement of the specific energy loss \dEdx in the detector gas. The tracks reconstructed in the ITS and the TPC are matched to hits in the other detectors inside the magnet located at larger radii. The Transition Radiation Detector (TRD)~\cite{Cortese:519145} surrounding the TPC provides hadron and electron identification via the measurement of the specific energy loss \dEdx and transition radiation. During the \PbPb (p--Pb) data taking period it covered 7/18 (13/18) of the full azimuth. Therefore, only in the \PbPb analysis it was used to verify the amount of hadron contamination within the electron identification strategy at low transverse momentum (see Section~\ref{anaPbPb}). The Time-Of-Flight array (TOF)~\cite{Cortese:2002kf}, based on Multi-gap Resistive Plate Chambers (MRPCs), provides hadron rejection at low transverse momentum via the time-of-flight measurement, within the electron identification strategy applied in both analyses. The T0 detectors, arrays of Cherenkov counters, located at $+$350~cm and $-$70~cm from the interaction point along the beam direction~\cite{Cortese:2004aa} provided, together with the TOF detector, the precise start time for the time-of-flight measurement in the \pPb analysis. For central \PbPb events the start time was estimated only using the particle arrival times at the TOF detector. 

The SPD, the T0 detectors as well as the V0 scintillator arrays, placed on both sides of the interaction point at 2.8 $ < \eta <$ 5.1 (V0-A) and $-$3.7 $ < \eta <$ $-$1.7 (V0-C), respectively, can be employed to define a minimum-bias trigger. The two Zero Degree Calorimeters (ZDC), that are symmetrically located 112.5~m from the interaction point on either side, were used in the offline event selection to reject beam-gas interactions by correlating the time information with the one from the V0 detectors.

The \PbPb and \pPb data presented here were recorded in 2010 and 2013, respectively.
Minimum-bias \pPb collisions were selected by requiring coincident signals in V0-A and V0-C (V0AND condition). Beam-gas interactions were rejected offline by the aforementioned correlation of the ZDC and V0 time information. The \PbPb collisions were collected with two different minimum-bias interaction triggers. The first trigger condition required signals in two of the following three detectors: SPD (two hits in the outer SPD layer), V0-A and V0-C. The second trigger condition required a coincidence between V0-A and V0-C. Both minimum-bias trigger conditions had efficiencies larger than 95\% for hadronic interactions, whereas the second rejected electromagnetic processes to a large extent~\cite{Abelev:2013qoq}. 
Only events with a primary vertex within $\pm$~10~cm from the centre of the detector along the beam direction were considered in the \pPb and \PbPb analyses. The \PbPb events were categorised into centrality classes by fitting the sum of the two V0 signal amplitudes with a geometrical Glauber-model simulation~\cite{Miller:2007ri}, as described in~\cite{Abelev:2013qoq}. The Glauber-model simulation yields a value of 18.93~$\pm$~0.74~$\rm mb^{-1}$ for the average nuclear overlap function $\langle T_{\rm AA} \rangle$ for the 20\% most central \PbPb collisions considered in the analysis. About 100 and 3 million \pPb and 20\% most central \PbPb events passed the offline selection criteria corresponding to an integrated luminosity of \mbox{$L_{\mathrm{int}}^{\rm pPb} = 47.8 \pm 1.6$ $\mu$b$^{-1}$} and \mbox{$L_{\mathrm{int}}^{\rm PbPb} = 2.2 \pm 0.2$ $\mu$b$^{-1}$}, respectively.

\section{Analysis overview and electrons from background sources}\label{Ana}

The identification of electrons from beauty-hadron decays is divided into the following steps:
\begin{itemize}
\item selection of tracks with good quality,
\item electron identification (eID),
\item determination of the electron yield from beauty-hadron decays.
\end{itemize}

The signal contains both electrons from direct decays ($\rm{b} \rightarrow \rm{e}$, branching ratio: $\approx$11\%) as well as cascade decays ($\rm{b} \rightarrow \rm{c} \rightarrow \rm{e}$, branching ratio: $\approx$10\%) of hadrons that contain a beauty (or anti-beauty) quark~\cite{Agashe:2014kda}.
Throughout the paper the term `electron' denotes both electron and positron. 
The track selection procedure is identical to previous analyses on the production of electrons from beauty-hadron decays~\cite{Abelev:2012sca,Abelev:2014gla}. The selection criteria are the same in the \pPb and \PbPb analyses, except for the restriction of the geometrical acceptance in rapidity, which was adjusted in each collision system to the region where the TPC could provide optimal electron identification, taking into account the detector and running conditions during each data-taking period. In \PbPb collisions this corresponds to the rapidity range $|y_{\rm lab}| \le 0.8 $ and in \pPb to $|y_{\rm lab}| \le 0.6 $. The tracks were required to have associated hits in both SPD layers, in order to minimise the contribution of electrons from photon conversions in the ITS detector material and the fraction of tracks with misassociated hits (see below).

The electrons were identified with the TPC and the TOF detectors via the measurement of their respective signal, specific energy loss in the gas (d$E$/d$x$) and the time-of-flight. The selection variable (hereafter $\nsigmaTPC$ or $\nsigmaTOF$) is defined as the deviation of the measured signal of a track with respect to the expected signal for an electron in units of the corresponding detector resolution ($\sigma_{\rm TPC}$ or $\sigma_{\rm TOF}$). The expected signal and the resolution originate from parametrisations of the TPC and TOF detector signals, described in detail in~\cite{Abelev:2014ffa}. For both analyses, particles were accepted with the TPC as electron candidates if they satisfied the condition $-0.5 < \nsigmaTPC < 3$. This asymmetric selection was chosen to remove hadrons, that are mainly found at negative \nsigmaTPC values. However, at low and high transverse momentum, the eID strategy based on TPC is subject to contamination from pions, kaons, protons and deuterons. To resolve these ambiguities, a selection cut of \mbox{$|\nsigmaTOF| \leq 3$} was applied for the whole \pt range in the \PbPb analysis and for \pt $\le$~2.5~\GeVc in the \pPb analysis. The remaining hadron contamination was determined via data-driven methods in the \pPb analysis and subtracted statistically (see Section~\ref{AnapPb}). The technique used for the \PbPb analysis is described in Section~\ref{anaPbPb}.

The electrons passing the track and eID selection criteria originate, besides from beauty-hadron decays, from the following background sources. In what follows, prompt and non-prompt contributions are marked in parentheses as `P' and `NP', respectively:
\begin{itemize}
\item (P) Dalitz and di-electron decays of prompt light neutral mesons ($\pi^0, \eta, \rho, \omega, \eta', \phi$), 
\item (P) di-electron decays of prompt heavy quarkonia (J/$\psi$, etc.).
\item (NP) decay chains of hadrons carrying a strange (or anti-strange) quark,
\item (NP) photon conversions in the detector material,
\item (NP) semi-leptonic decays of prompt hadrons carrying a charm (or anti-charm) quark.
\end{itemize}

The measurement of the production of electrons from beauty-hadron decays exploits their larger mean proper decay length $c\tau \approx$~500~$\mu$m~\cite{Agashe:2014kda}) compared to that of charm hadrons and most other background sources, resulting in a larger average impact parameter.
The sign of the impact parameter value is attributed based on the relative position of the track and the primary vertex, i.e.\ if the primary vertex lies on the left- or right-hand side of the track with respect to the particle momentum direction in the transverse plane.

For the presented analyses, the impact parameter was multiplied with the sign of the particle charge and of the magnetic field component along the beam axis (plus or minus for the two field orientations). With this definition, the sign of the impact parameter depends on whether the primary vertex lies inside or outside of the circle defined by the track projection in the transverse plane. Electrons from the conversion of photons in the detector material have an initial momentum with a very small angle to the direction of the photon. The magnetic field bends the track away from the primary vertex. Thus, they typically have an impact parameter \dmin~$< 0$. The asymmetric shape helps to differentiate this background source. It is important to include the field configuration, because the magnetic field direction was reversed during the \PbPb data taking period, which motivated this redefinition.

\begin{figure}[tb]
\centering
 \includegraphics[scale=0.38]{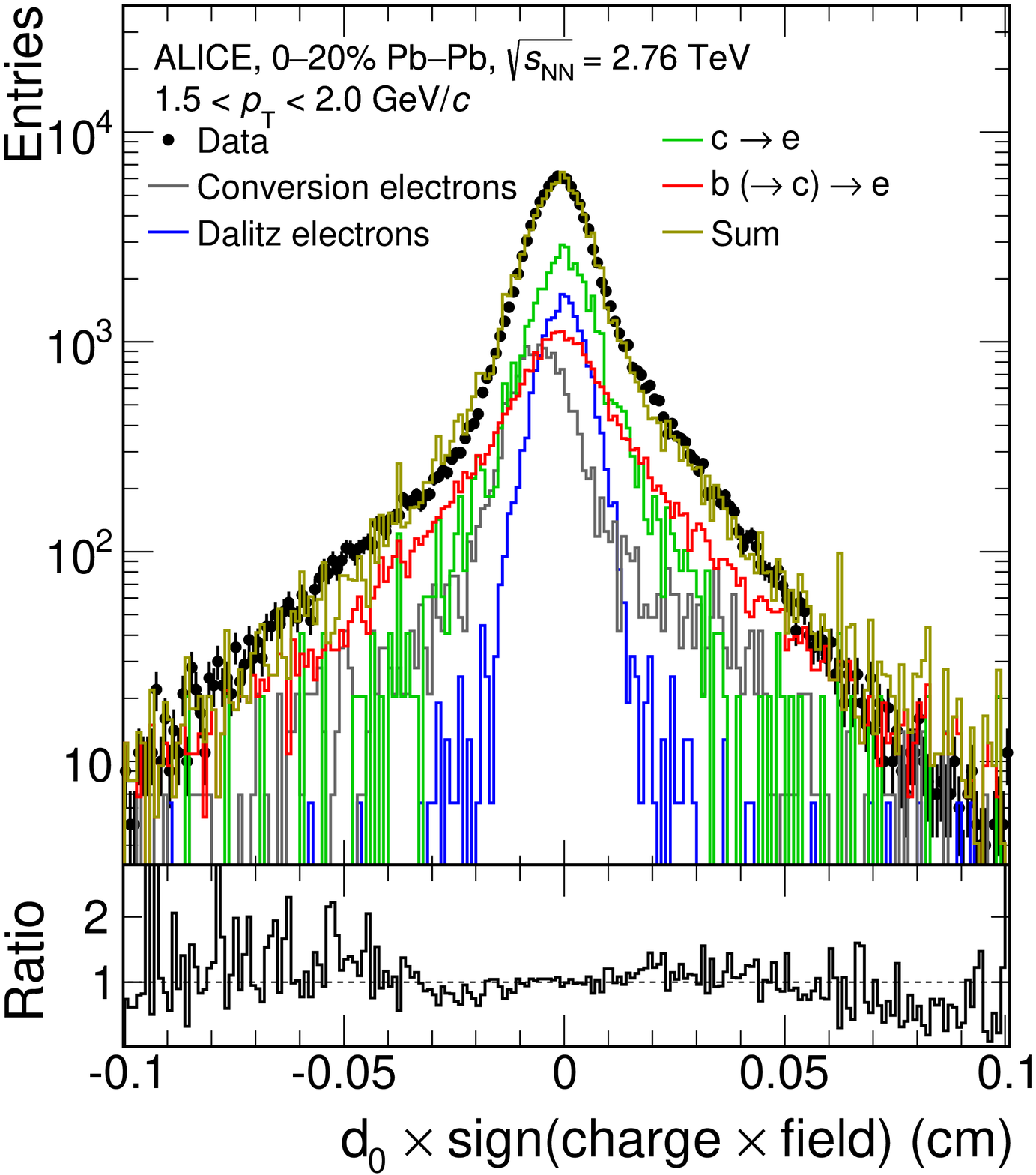}
  \hspace{\fill}
 \includegraphics[scale=0.38]{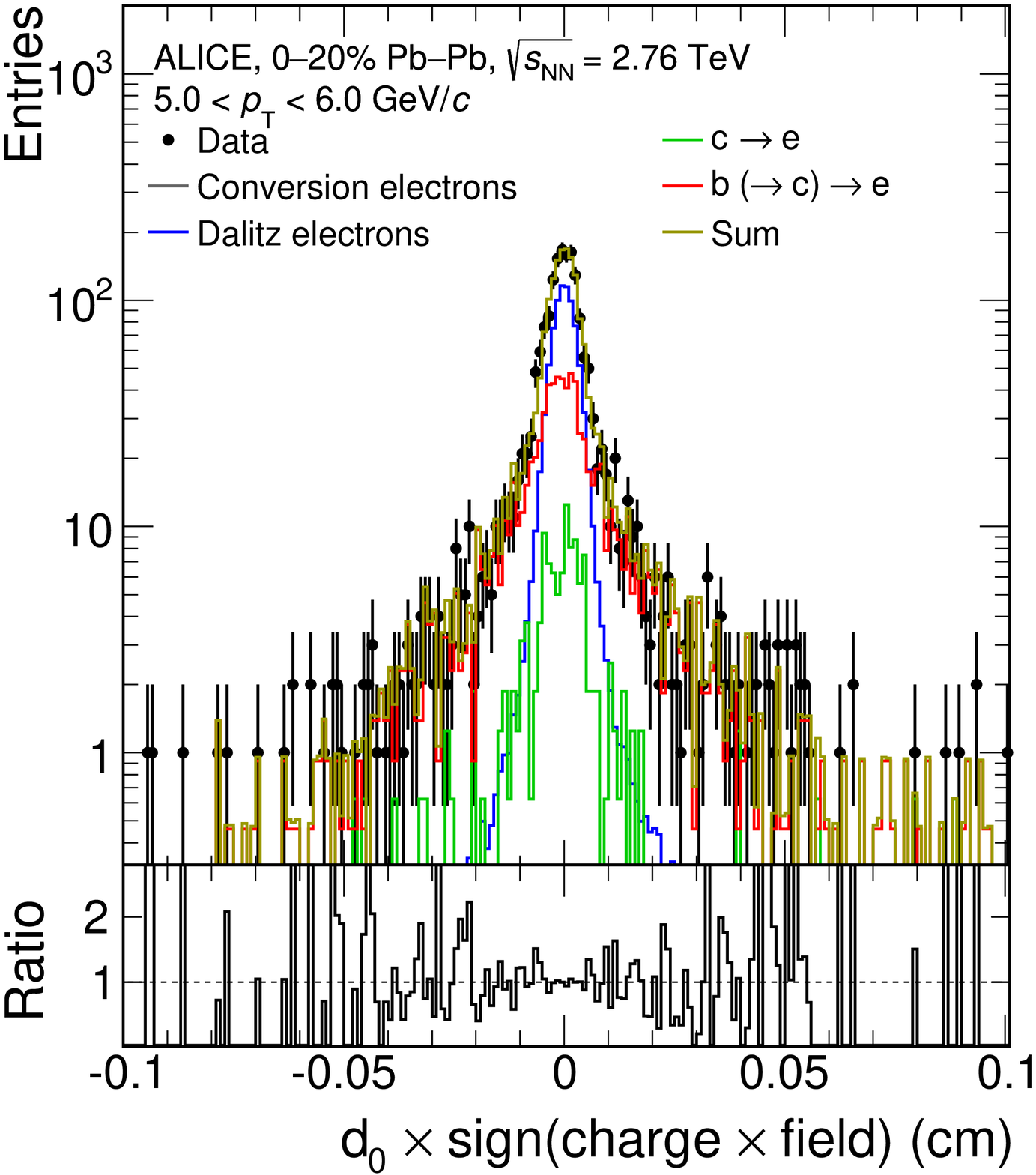}
 \caption{Impact parameter distribution for the interval (left) \ptrange{1.5}{2.0} and (right) \ptrange{5}{6} in the 20\% most central \PbPb collisions. The impact parameter value of each track was multiplied by the sign of the charge of each track and the sign of the magnetic field. The individual distributions for electrons from beauty-hadron and charm-hadron decays, from Dalitz-decays of light mesons, and from photon conversions were obtained by HIJING and PYTHIA simulations. The bottom panel shows the ratio of the data and `Sum'.}\label{fig::dca}
\end{figure}

Figure~\ref{fig::dca} shows for two \pt intervals the resulting distribution of the measured impact parameter value multiplied by the sign of the charge of each track and the sign of the magnetic field for the \pt interval \ptrange{1.5}{2.0} in the 20\% most central \PbPb collisions. The impact parameter distributions for electrons from beauty- and charm-hadron decays, from Dalitz decays of light mesons, and from photon conversions are also drawn for comparison. The distributions were obtained from Monte Carlo simulations and normalised to the data using the fit values described in Section~\ref{anaPbPb}. The distribution for electrons from photon conversions is, as explained before, visible as an asymmetric and shifted distribution. The impact parameter distribution of electrons from prompt sources, such as Dalitz and quarkonium decays, is determined by the impact parameter resolution. The electrons from these sources are thus categorised as Dalitz decays within both analyses. 

The Monte Carlo simulations were produced as follows. A sample of minimum-bias \PbPb collisions at \sqAA was generated with HIJING v1.36~\cite{Gyulassy:1994ew} for efficiency and acceptance corrections as well as to obtain the impact parameter distributions for photon conversions and Dalitz decays. To increase the statistics of electrons from charm- and beauty-hadron decays, a signal enhanced sample was generated using pp events produced by the generator PYTHIA v6.4.21~\cite{Sjostrand:2006za} with Perugia-0 tune~\cite{Skands:2010ak}. Each added pp event contains one $\rm c\overline{c}$ or $\rm b\overline{b}$ pair. For the \pPb analysis, the same procedure was used.
The generated particles were propagated through the ALICE apparatus using GEANT3~\cite{Brun:1994aa} and a realistic detector response was applied to reproduce the performance of the detector system during data taking.

The inclusive yield of electrons originating from strange-hadron decays is small compared to the other background sources. However, as these electrons originate from secondary $\pi^0$ from strange-hadron decays (\KS, \KL, $\rm{K}^{\pm}$, $\Lambda$) and three prong decays of strange hadrons (\KL,$\rm{K}^{\pm}$), the impact parameter distribution is broader than that of electrons from Dalitz and di-electron decays of other light neutral mesons. Sections~\ref{AnapPb} and~\ref{anaPbPb} describe how the analyses handle this background contribution.

Although requiring hits in both SPD layers, electrons from photon conversions in detector material with production radii outside the SPD layers were observed to have passed the track selection. These electron tracks are wrongly associated with signals of other particles in the inner detector layers. Within this paper these electrons are called `mismatched conversions'. The amount of mismatched conversions depends on the track multiplicity within the event and thus has a larger impact for the \PbPb analysis. Sections~\ref{AnapPb} and~\ref{anaPbPb} outline how the analyses deal with the mismatched conversions.

The impact parameter distributions of electrons from most background sources are narrow compared to the one of electrons from beauty-hadron decays. By applying a minimum cut on the absolute value of the impact parameter \dminabs, the fraction of electrons from beauty-hadron decays can thus be enhanced. The remaining background can be described using a cocktail method and subtracted statistically to obtain electrons from beauty-hadron decays~\cite{Abelev:2012sca,Abelev:2014gla}. This method was applied in the \pPb analysis and is described in detail in Section~\ref{AnapPb}. Another technique, used in the \PbPb analysis (see Section~\ref{anaPbPb}), is to make use of the whole impact parameter distribution, i.e.\ to compare the impact parameter distributions of the various electron sources from simulation (templates) with the impact parameter distribution of all measured electron candidates to estimate the individual contributions.

\section{Data analysis in \pPb collisions}\label{AnapPb}

The identification of electrons from beauty-hadron decays in the \pPb analysis is based on the selection of electrons with large impact parameters. This method was already applied in \pp collisions at \s~=~2.76~TeV and \s~=~7~TeV~\cite{Abelev:2014gla,Abelev:2012sca}. Since the impact parameter distribution of electrons from beauty-hadron decays is broader compared to the one of electrons from most background sources~(see Section~\ref{Ana}), the requirement of a minimum absolute impact parameter enhances the signal-to-background \mbox{(S/B)} ratio of electrons from beauty-hadron decays. The remaining background due to hadron contamination and electrons from background sources was obtained via a data-driven method and from Monte Carlo simulations re-weighted to match the \pt distributions of the background sources in data, respectively, and then subtracted.

\subsection{Extraction of electrons from beauty-hadron decays}

Electron candidates with an impact parameter \dminabs $>0.0054+0.078 \times \rm{exp}(-0.56 \times \textit{p}_{\rm T})$ (with \dmin in cm and \pt in \GeVc) were selected. This selection criterion was determined from Monte Carlo simulations to maximise the significance for electrons from beauty-hadron decays. The selection of the minimum impact parameter is \pt dependent, because the width of the impact parameter distribution, the S/B ratio as well as the true impact parameter distribution of the various electron sources~\cite{Abelev:2012sca} are \pt dependent.

\begin{figure}[tb]
\centering
  \includegraphics[scale=0.38]{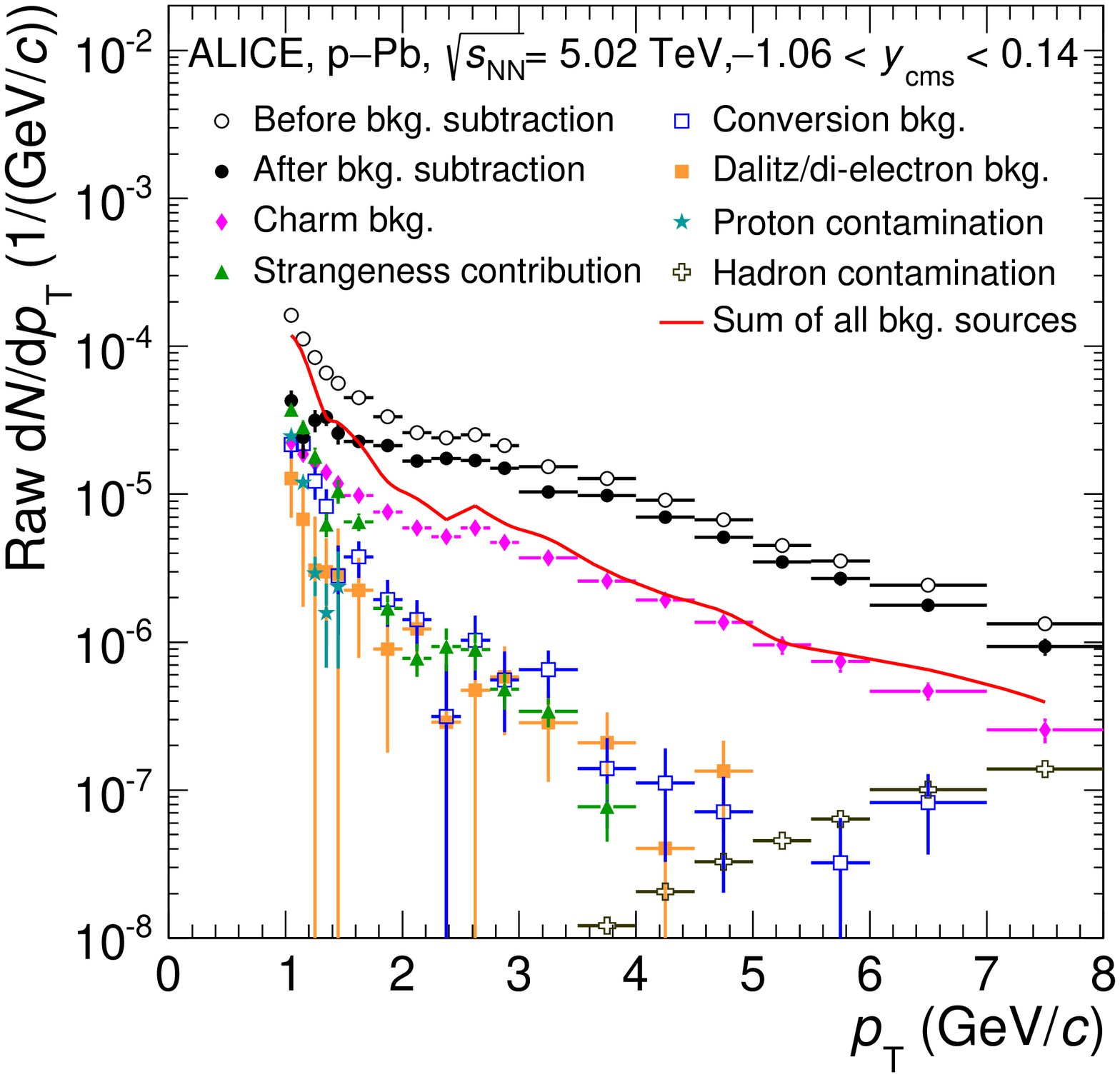}
  \hspace{\fill}
  \includegraphics[scale=0.39]{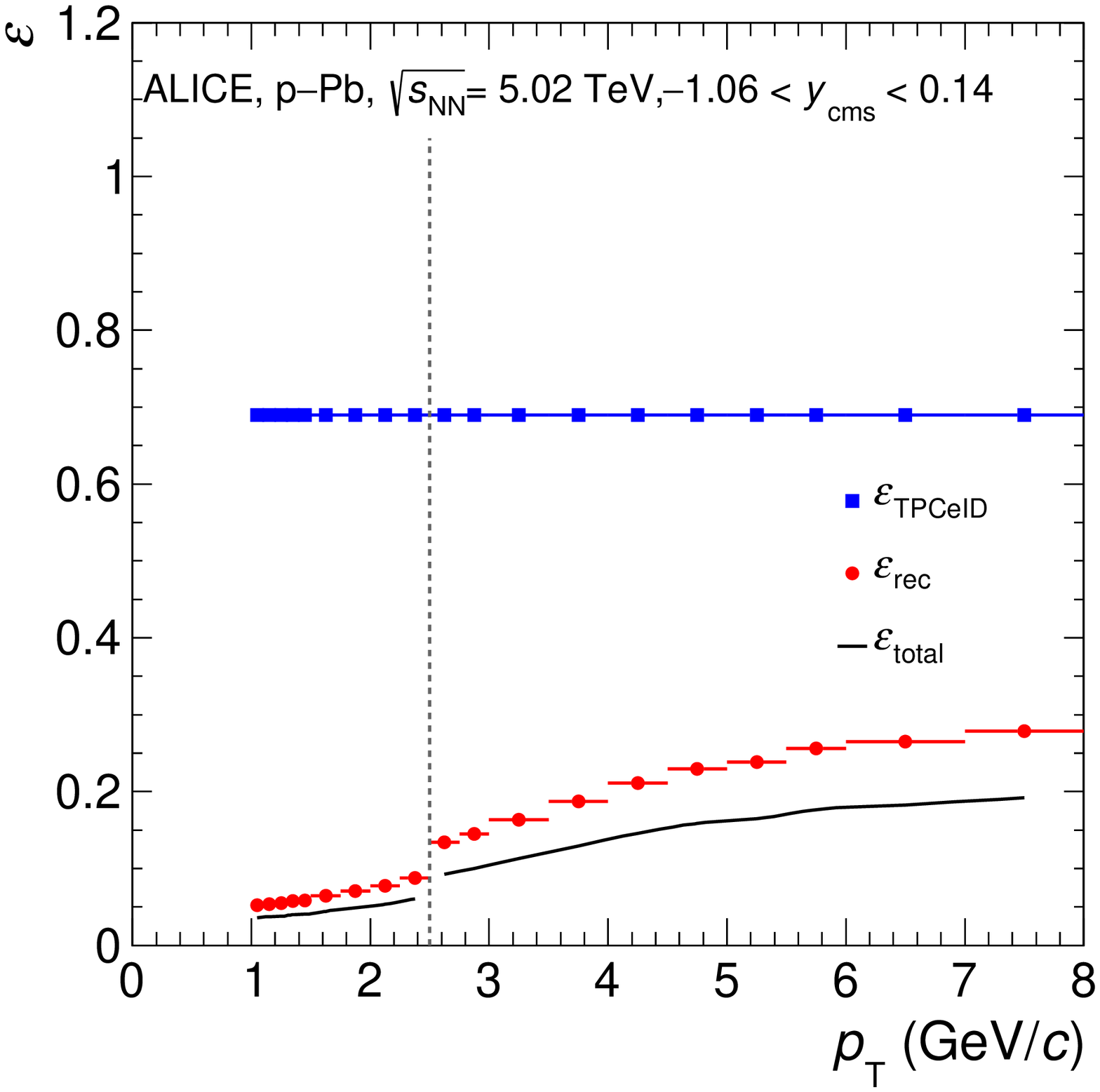}
  \caption{(left) Raw transverse momentum distribution of electrons after track, eID and impact parameter requirement in comparison with the proton and hadron contamination as well as electrons from the different background sources in \pPb collisions. The contributions of electrons from strange-hadron decays are included in the distributions labelled `Dalitz/di-electron bkg.' and `Conversion bkg.'. The error bars represent the statistical uncertainties. (right) Efficiencies for the \pPb analysis as a function of transverse momentum (see text and Equation~\ref{eq:cross_section} for details). The vertical dashed line indicates the switch of the eID between the TPC and TOF and TOF-only method.
}
\label{fig::raw_pPb}
 \end{figure}

The number of hadrons passing the track selection, eID, and the minimum impact parameter requirement was estimated at high transverse momentum (\pt $\geq 4~\GeVc$) by parametrising the TPC \nsigmaTPC distribution in momentum slices, and it was subtracted~\cite{Abelev:2012xe}. Above a \pt of $4~\GeVc$, the hadron contamination increases with transverse momentum and reaches 10\% at 8~\GeVc, see Fig.~\ref{fig::raw_pPb} (left). At low transverse momentum (\pt $\leq 4~\GeVc$), the hadron contamination is negligible except in the transverse momentum interval \ptrange{1}{1.2}, see Fig.~\ref{fig::raw_pPb} (left), where electrons cannot be distinguished from protons via the measurement of specific energy loss in the TPC gas. In addition, the requirement of a minimum impact parameter increases the relative contribution of secondary protons originating from e.g.\ $\Lambda$ and $\Sigma^{+}$ decays, which have larger impact parameter values compared to electrons from beauty-hadron decays. The relative abundance of protons in the electron candidate sample was determined by using the ITS particle identification capabilities, because electrons and protons can be separated with ITS in this momentum interval. The ITS energy loss signal was fitted with data-derived templates for electrons and protons. The templates were obtained in \pt-bins by selecting electrons and protons with tight selection criteria in TOF and TPC. The estimated proton contribution, which is $\approx$10\% (4\%) in the \pt interval \ptrange{1}{1.1} (\ptrange{1.1}{1.2}), was subtracted statistically from the measured electron candidate \pt distribution.

Figure~\ref{fig::raw_pPb} (left) shows the transverse momentum distribution of electrons passing the track, eID, and impact parameter selection, before efficiency corrections. The contributions due to the proton and hadron contamination at low and high \pt, respectively, determined via the aforementioned methods are shown. Also shown are the distributions of electrons originating from the various background sources, which were obtained using the Monte Carlo simulations described in Section~\ref{Ana}. To match the measured shapes, the \pt differential yields of the background sources were re-weighted in the Monte Carlo simulations prior to the propagation through the ALICE apparatus with GEANT3. As there is no measurement of the $\pi^0$ production cross section in \pPb collisions available, the $\pi^0$ input was based on the measured charged-pion spectra~\cite{Abelev:2013haa,Adam:2015pionshighpt} assuming $N_{\pi^0} = (N_{\pi^+} + N_{\pi^-})/2$. Due to the requirement of a minimum impact parameter, the contribution of electrons from decays of secondary $\pi^{0}$ from strange-hadron decays is comparable with the one from primary decays. Therefore the measured \pt spectra of $\rm{K}^{\pm}$, \KS and $\Lambda$~\cite{Abelev:2013haa} were used to compute the corresponding weights. To obtain the weights, the pion and strange-hadron spectra were parameterised with a Tsallis function as described in~\cite{Abelev:2012xe}. The contribution of electrons originating from secondary pions from strange-hadron decays or three-body decays of strange hadrons is shown in Fig.~\ref{fig::raw_pPb} (left). The other light mesons ($\eta$, $\rho$, $\omega$, $\eta'$ and $\phi$), which contribute little, via Dalitz decays and photon conversions compared to primary $\pi^{0}$ decays, were re-weighted via \mt-scaling of the $\pi^{0}$ spectrum~\cite{Abelev:2012xe}. The electron background from charm-hadron decays was estimated based on the $\rm{D}^{0}$,~$\rm{D}^{+}$ and $\rm{D}_{s}^+$ meson production cross section measurements with ALICE~\cite{Abelev:2014hha} in the transverse momentum intervals \ptrange{1}{16}, \ptrange{2}{24} and \ptrange{2}{12}, respectively. In a first step the measurements were extrapolated to the \pt interval \ptrange{1}{24} by assuming constant ratios $\rm{D}^{0}$/$\rm{D}^{+}$ and $\rm{D}_{s}^{+}$/$\rm{D}^{0}$ from the measured D meson production cross sections. Next the \pt differential production cross sections were extrapolated to \pt~=~50~\GeVc via FONLL pQCD calculations. About 10\% of the electrons with \pt $\le$~8~\GeVc originate from the extrapolated D meson high-\pt region (\pt~$\geq 24$~\GeVc). The electron contribution from $\Lambda_{\rm c}^{+}$ decays was estimated using the ratio $\sigma \left( \Lambda_{\rm c}^{+} \right)$/$\sigma \left( \rm{D}^{0} + \rm{D}^{+} \right)$ measured by the ZEUS Collaboration~\cite{Abramowicz:2013eja}. Analogous to the light mesons, the measured D meson \pt spectra were also used to re-weight the \pt distributions in the Monte Carlo simulations.

The signal of electrons from beauty-hadron decays was obtained after subtraction of the aforementioned background contributions from the measured electron candidate sample after track selection, eID and impact parameter requirement. The resulting \pt spectrum is shown in Fig.~\ref{fig::raw_pPb} (left). At $\pt = 1~\GeVc$, the number of electrons from beauty-hadron decays is approximately equal to the one from charm-hadron decays, from Dalitz decays of light mesons, from strange-hadron decays and from photon conversions, resulting in a S/B ratio of approximately $1/3$. With increasing \pt the background electron yield from Dalitz decays of light mesons, from strange-hadron decays and from photon conversions quickly decreases compared to the contribution of electrons from charm-hadron decays. In the \pt interval \ptrange{4.5}{5}, the S/B ratio reaches its maximum of 3. Here the electron background mostly originates from charm-hadron decays. At higher \pt, the S/B ratio decreases again due to the increasing hadron contamination. 
Other background sources, such as di-lepton decays of J/$\psi$ mesons are negligible due to the minimum impact parameter selection. The yield of electrons from Drell-Yan processes is negligible over the whole \pt range~\cite{Abelev:2012xe}.
 
The raw yield of electrons from beauty-hadron decays $N_{\rm e, raw}$ was then corrected for the geometrical acceptance and for the efficiency ($\epsilon_{\rm rec}$) of the track reconstruction, matching and selection criteria, TOF electron identification and minimum impact parameter requirement using the Monte Carlo simulations. The efficiency of the TPC electron identification selection ($\epsilon_{\rm TPC eID}$) was determined to be 69\% via a data-driven approach based on the \nsigmaTPC distributions~\cite{Abelev:2012xe}. The transverse-momentum dependence of the efficiencies is shown in Fig.~\ref{fig::raw_pPb} (right). The total efficiency shows a significant \pt dependence, mainly due to the $d_0$ cut. The effects of the finite momentum resolution and the energy loss due to Bremsstrahlung were taken into account in a bin-by-bin \pt resolution correction step based on a Monte Carlo simulation~\cite{CowanStatistical,Abelev:2012sca}. 

The $\pt$-differential invariant cross section of electrons from beauty-hadron decays, $\rm (e^+ + e^-)/2$, is thus given as:
\begin{equation}
\frac{1}{2\pi\pt}\frac{{\rm d^2}\rm \sigma}{{\rm d}\pt{\rm d}y}=
\frac{1}{2}
\frac{1}{2\pi \pt^{\rm{centre}}} 
\frac{1}{\Delta y \Delta \pt}
\frac{N_{\rm e, raw}}{\epsilon_{\rm rec} \times \epsilon_{\rm TPC eID}}
\frac{\smbeq}{N_{\rm mb}},
\label{eq:cross_section}
\end{equation}
where $\pt^{\rm{centre}}$ is the centre of the \pt\ bin with width $\Delta\pt$ and $\Delta y$ denotes the geometrical acceptance in $|y_{\rm lab}|$ to which the analysis was restricted. $N_{\rm mb}$ is the total number of analysed minimum-bias events. The \pPb cross section for the minimum-bias V0 trigger condition, which has an efficiency of more than 99\% for non-single-diffractive (NSD) \pPb collisions~\cite{ALICE:2012xs}, is $\smbeq = 2.09 \pm 0.07$~b~\cite{Abelev:2014epa}.

\subsection{Systematic uncertainties estimation}\label{pPbsys}

An overview of the relative systematic uncertainties is shown in Table~\ref{tab::syst}. The systematic uncertainties were estimated as a function of \pt by repeating the analysis with modified track selection and eID criteria and by varying the background yields within their estimated uncertainties. 

The uncertainty of the tracking results from differences in data and Monte Carlo simulations for the track reconstruction with the ITS and the TPC, which includes the uncertainty of finding a hit in the ITS for a track reconstructed in the TPC. The latter uncertainty (3\%) was taken from~\cite{Abelev:2014dsa}, where the effect was studied for charged particles. The TOF-TPC matching uncertainty (5\%) was obtained by comparing the matching efficiency of electrons from photon conversions identified via topological selections in data and Monte Carlo simulations. The TOF eID uncertainty was derived by repeating the analysis with different eID selection criteria. At high \pt the TOF was not used in the analysis and thus the corresponding uncertainty does not apply in this region. The uncertainty of the TPC eID was estimated in the same way as for the TOF eID. The systematic uncertainty of the determination of the hadron contamination ranges from 1\% to 6\%, i.e. increasing as the contamination itself with increasing \pt.

The systematic uncertainty of the minimum impact parameter requirement was evaluated by varying this selection criterion by $\pm$1 $\sigma$, where $\sigma$ corresponds to the measured impact parameter resolution~\cite{Abelev:2012sca}. At 1~\GeVc (8~\GeVc) this corresponds to a $\approx$10\% ($\approx25$\%) variation of the cut value. 

The number of electrons from photon conversions increases quickly with decreasing transverse momentum (see Fig.~\ref{fig::raw_pPb}, left). The difference in yield of mismatched conversions in data and Monte Carlo simulations was estimated and assigned as a systematic uncertainty. For this purpose pions from \KS decays identified via topological and invariant mass cuts~\cite{PhysRevLett.111.222301} can be used, because their decay vertex can be reconstructed, in contrast to electrons from photon conversions, for which it is more difficult due to their small opening angle. 
The yield of pions from \KS decays was studied as a function of the production vertex with and without requiring a signal in both SPD layers and compared with the corresponding results from Monte Carlo simulations. The difference in yield was propagated into the simulation by renormalising the number of electrons from photon conversions. Repeating the full analysis with the varied conversion yield results in the uncertainties listed in Table~\ref{tab::syst}. 

The systematic uncertainty arising from the subtraction of electrons from the various background sources was evaluated by propagating the statistical and systematic uncertainties of the light-meson, strange- and charm-hadron measurements used as input to re-weight the \pt distributions in Monte Carlo simulations. Uncertainties due to the \mt-scaling of the background yields, estimated as 30\%~\cite{Abelev:2012xe}, and the extrapolation of the D meson \pt distributions to the unmeasured transverse momentum regions were included. The latter was obtained by using the uncertainties of the various D meson ratios and by using a power-law fit instead of FONLL pQCD calculations for the extrapolation of the \pt reach to 50~\GeVc. The uncertainty of the contribution of electrons from $\Lambda_{\rm c}$ decays was estimated by varying the ratio $\sigma \left( \Lambda_{\rm c} \right)$/$\sigma \left( \rm{D}^{0} + \rm{D}^{+} \right)$ by $\pm$~50\% of its original value. The resulting uncertainty is negligible compared to the overall systematics, because the $\Lambda_{\rm c}$ contribution is small, less than 10\%.

Over the whole \pt range, the systematic uncertainty due to the subtraction of electrons from charm-hadron decays dominates. The uncertainty due to the subtraction of electrons from light-hadron decays is large at very low \pt, but decreases quickly with increasing \pt as does the overall yield of this background source, as shown in Fig.~\ref{fig::raw_pPb} (left). At high \pt, the uncertainty of the hadron contamination increases. 

The influence due to the form factor of electrons from charm and beauty hadron decays
as well as light neutral mesons was studied using different decayers and estimated to be negligible.

As the individual sources of systematic uncertainties are uncorrelated, they were added in quadrature to obtain the total systematic uncertainty. The total uncertainty amounts to 38\% for the lowest \pt interval and decreases to 12\% at \pt~=~8~\GeVc.

The systematic uncertainty due to the determination of the nucleon--nucleon cross section for the minimum-bias trigger condition is 3.7\%~\cite{Abelev:2014epa}.

\begin{table}[h]
\centering
\scalebox{0.9}{
\begin{tabular}{|l|c|c|c|}

\hline
Source      & \ptrange{1}{2.5}& \ptrange{2.5}{8} \\
\hline
Tracking and matching   &   ~5.6\%          &      ~5.2\%           \\
TOF matching and eID    &   ~5.4\%         &    n/a             \\
TPC eID                 & 3\%           &  3\%            \\
Hadron contamination    & n/a              &   1\% to 6\%  \\
\hline
Minimum \dmin requirement  &            5$\%$                   &      5$\%$                \\
Mismatched conversions  & 4\% to 0.3\%     & negligible                \\
Light- and strange-hadron decay bkg. & 17\% to 1.5\% & 1.3\% to     0\%       	     \\
Charm-hadron decay bkg. & 32\% to 9.6\%  &     8.9\% to 6.2\%               \\
\hline
\hline
Total                   & 38\% to 14\% &     12\%               \\
Normalisation uncertainty &  \multicolumn{2}{|c|}{3.7\%}   \\
\hline
\end{tabular}
}
\caption{Systematic uncertainties in the \pPb analysis. The two columns with the different momentum intervals correspond to the TPC and TOF and TPC-only eID strategies. Individual sources of systematic uncertainties are \pt dependent, which is reported using ranges. The lower and upper values of the interval, respectively, represent the uncertainty at \pt~=~1~\GeVc (\pt~=~2.5~\GeVc) and \pt~=~2.5~\GeVc (\pt~=~8~\GeVc) for the TPC and TOF (TPC-only) eID strategy. The lower and upper values of the interval for the hadron contamination are \pt~=~4~\GeVc and 8~\GeVc. The second group of entries in the table is related to the method used to extract the electrons from beauty-hadron decays.}
\label{tab::syst}
\end{table}

\section{Data analysis in \PbPb collisions}\label{anaPbPb}
In the \PbPb analysis, the yield of electrons from beauty-hadron decays was extracted using the full information contained within the impact parameter distribution of all electron candidates. From the shape of the impact parameter distribution within one \pt interval, it is possible to infer the contributions from the different electron sources (see Section~\ref{Ana}). Templates for these distributions were obtained from Monte Carlo simulations including effects such as particle lifetime and the detector response. The templates were then added with appropriate weights to reproduce the measured impact parameter distribution for all electron candidates. Examples are shown in Fig.~\ref{fig::dca}. The template fits were performed using the method proposed in~\cite{Barlow1993219}. 
The approach relies on the accurate description of the impact parameter distributions in Monte Carlo simulations. Thus, detailed studies of differences between the impact parameter distributions in data and Monte Carlo simulations were performed. Differences were corrected for, while the related uncertainties were propagated as detailed below. For the template fit method, four classes of electron sources were distinguished. Their impact parameter distributions, as provided by the Monte Carlo simulations for each \pt interval, will be referred to as fit templates in the following. The four categories correspond to electrons from beauty-hadron decays, from charm-hadron decays, from photon conversions and electrons from other processes, which will be referred to as `Dalitz electrons'. The latter is dominated by electrons from Dalitz decays of light neutral mesons. Given that these electrons essentially originate from the interaction point with respect to the detector resolution, the measured impact parameter distribution depends only on the transverse momentum of the electron. Similarly, the remaining hadron contamination consists of particles mostly produced close to the interaction point making its impact parameter distribution similar to that of the Dalitz electrons.

\begin{figure}[tb]
\centering
   \includegraphics[scale=0.4]{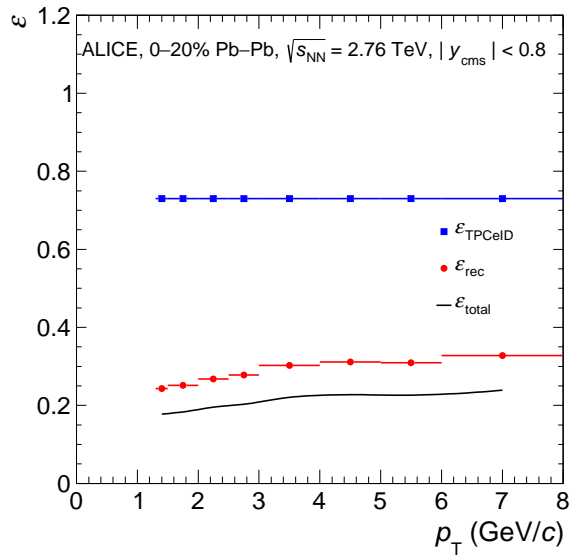}
  \caption{Efficiencies of the different track selection steps for the measurement in central Pb--Pb collisions.}
\label{fig::eff_PbPb}
 \end{figure}

\subsection{Extraction of electrons from beauty-hadron decays} \label{MethodPbPb}
The fit templates from the Monte Carlo simulations can be considered as random samples of the unknown true distributions. For each of the four electron sources considered in the previous section, there is a number of counts in the template for each impact parameter bin (see Fig.~\ref{fig::dca}). The number of counts from a particular electron source $j$ in a particular bin $i$ is called $a_{ji}$. Its unknown expectation value is called $A_{ji}$ and is considered as a free parameter of the fit. The fit function is the sum of the expectation values, each weighted with the appropriate amplitude parameter $p_{j}$: $f_{\textrm{i}} = \sum_{j} p_{j} ~ A_{ji}$. The bin counts of the impact parameter templates are connected to their expectation values via Poisson statistics. The same relation holds between the fit function and the data ($d_{i}$) within each impact parameter bin leading to the likelihood distribution~\cite{Barlow1993219}
\begin{equation}
\log \mathcal{L} = \sum_{i} d_{i} \log f_{i} - f_{i} + \sum_{j} \sum_{i} a_{ji} \log A_{ji} - A_{ji} ~. \label{LikelihoodFormula}
\end{equation}
This gives one free amplitude parameter for each electron source ($p_{j}$) and one free expectation value parameter for each electron source and impact parameter bin ($A_{ji}$). The main parameters of interest are the $p_j$, in particular $p_{\mathrm{beauty}}$, while the nuisance parameters $A_{ji}$ arise due to the finite statistics of the templates. Evaluating the full likelihood distribution in several hundred dimensions is challenging. Therefore a simpler approach is to use the maximum likelihood as an estimator for the amplitudes of the electron sources.

An iterative procedure to find the maximum likelihood with respect to the $A_{ji}$ for fixed $p_{j}$ is suggested in~\cite{Barlow1993219}. Numerical minimisation is then performed only for the $p_{j}$. Equations for the iterative procedure can be found by setting the differentials $\mathrm{d} \mathcal{L}/ \mathrm{d} A_{ji}$ to zero. Solving these equations for $A_{ji}$, yields an iterative rule for each bin.

For a bin $i$ with a finite number of entries from data, but zero counts in any of the templates, the likelihood distribution of the $A_{ji}$ is not well represented by its maximum. This happens mostly in the tails of the distributions (see Fig.~\ref{fig::dca}), where the contribution of electrons from beauty-hadron decays dominates. Thus, for this case only the contribution from this source was considered.

To obtain the raw yield of the signal, i.e. electrons from beauty-hadron decays, in a given \pt interval, 
the number of electrons in the template was scaled by the amplitude parameter $p_{\mathrm{beauty}}$. 
As in the \pPb analysis, the raw yield was then corrected for the geometrical acceptance, the track reconstruction and selection criteria, the TOF acceptance, and the TOF eID using 
Monte Carlo simulations. The TPC eID efficiency ($\epsilon_{\rm TPC eID}$) was determined via a data-driven approach using 
electrons from photon conversions identified via topological cuts and the 
invariant mass~\cite{Abelev:2014ypa}. The corresponding \nsigmaTPC distributions were fitted with the 
function ${Landau} \cdot Exp \otimes Gauss$~\cite{Abelev:2012xe}, which describes the distributions 
including fluctuations, and the efficiency determined as the ratio of electrons before and after the 
TPC eID selection criterion (see Section~\ref{Ana}). 
Next the \pt spectrum was unfolded. The off-diagonal elements of the response matrix are small. For this reason no regularisation was used in the unfolding procedure to avoid additional systematic uncertainties. The unfolding was done using a matrix inversion of the response matrix~\cite{CowanStatistical}. Due to the restricted \pt range of the measurement there is some dependence of the unfolded values on bins that have not been measured, mainly the adjacent bins. To solve this, the yield was measured in two further bins (\ptrange{1.1}{1.3} and \ptrange{8}{12}) and used only in the unfolding calculations. The statistical uncertainties were propagated accordingly.

To validate this signal extraction method, the template fit method was also applied to the \pPb data, where results were found to be consistent with the cut method described in Section \ref{AnapPb}.

\subsection{Systematic uncertainties estimation}\label{MethodUncertPbPb}

The systematic uncertainties are summarized in Table~\ref{tab::systPbPb}. They were estimated using data-driven methods where possible. An overview of the efficiencies of the different track selection steps may be found in Fig.~\ref{fig::eff_PbPb}.

The efficiency due to the ITS track selection criteria (hits in both SPD layers) does not depend strongly on the particle species. Thus, charged tracks could be used as a representative sample with respect to the geometric effects, such as inactive areas of the detector. The normalisation for the efficiency was performed by making use of phase space (pseudorapidity and azimuthal angle) regions where the efficiency was close to unity. Averaging over the phase space yields a proxy for the total efficiency which was compared between data and Monte Carlo simulations and yielded a difference of $2\%$. The uncertainty for non-geometric effects was estimated to be smaller than $3\%$.
The efficiencies of the requirements on charged tracks with good quality, the TOF matching and TOF eID depend more strongly on the particle type. Therefore, only an electron sample could be representative. It was obtained by selecting electrons from photon conversions. Due to the large particle multiplicity in central \PbPb collisions (resulting in a sizeable hadron contamination), the comparison was done using weak additional particle identification ($-1.5 < \nsigmaTPC < 4$), in more peripheral collisions ($20\rm{-}40\%,\, 40\rm{-}80\%$), and with different ITS track selection criteria (excluding signals in the innermost layer). To account for biases due to these additional criteria, they were varied and the results were checked for consistency. The estimated systematic uncertainties are about $3\%$ for the requirement of charged tracks with good quality and about $10\%$ for the TOF matching and eID.
The systematic uncertainty of the TPC eID includes differences in the eID efficiency for electrons from beauty-hadron decays and for electrons from photon conversion (due to the different pseudorapidity distributions) in the sample as well as the uncertainty of the extrapolation towards lower \nsigmaTPC. The uncertainty due to the modelling of the \nsigmaTPC distribution was checked by comparing different model descriptions with the standard one and by comparing with a sample of pions selected with the TRD and TOF. The total uncertainty of $5\%$ for the TPC eID is the quadratic sum of the following contributions: $2\%$ from the extrapolation, $2\%$ from the pseudorapidity dependence, $3\%$ from a possible \pt dependence and $2\%$ from the tail of the \nsigmaTPC distribution.

To estimate the statistical and systematic uncertainty on the extracted signal yield due to the maximum-likelihood fit, a Monte Carlo closure test was used. For this purpose, the templates were slightly smoothed and the result sampled with the statistics present in the measurement. The pseudo-data was created by using the measured contributions as input. The application of the template fit allowed for a comparison of the measured and true value. Repetitions of this process gave an estimation of the statistical and systematic contribution to the uncertainty. The charm yield of the test was varied to avoid underestimating the uncertainty in \pt intervals with downward fluctuations of the measured charm yield. The systematic uncertainty varies between 19\% and 6\% between the different \pt intervals.

There is an uncertainty in how well the impact parameter distributions of the different electron sources are described by the Monte Carlo simulations. Where possible, any differences were corrected for. The remaining uncertainty was propagated to the measured spectrum of electrons from beauty-hadron decays by changing the fit templates within their uncertainties.

The different resolution of the impact parameter (\dmin) with the given track and event selection criteria in Monte Carlo simulations and data was corrected for. The size of the correction was estimated by comparing the impact parameter distributions of primary pions, yielding a 10--12\% worse resolution in data compared to the Monte Carlo simulations in the \pt range of the measurement. To correct for this effect, a Gaussian distributed random number was added to each impact parameter value such that the resolution in the Monte Carlo simulations matched that of the data. The central values of the yield of electrons from beauty-hadron decays were estimated using a resolution correction of $10\%$. The yield using a correction of $12\%$ instead, differs by about $10\%$ at \pt~=~1.3~\GeVc with the difference decreasing quickly towards higher \pt. The effect of the correction was found to be negligible for the \pPb analysis.

Despite the strong eID requirements, there is a significant contamination of the electron sample by hadrons (mostly charged pions). The contribution was estimated using a clean TPC energy loss signal of pions identified with the TRD, which was fitted to the \nsigmaTPC distribution, suggesting a contamination of the electron candidate sample of about $15\%$ even for low transverse momentum. The contamination was not explicitly subtracted. The impact parameter distribution of charged hadrons is similar to that of the Dalitz template. This means that the contribution of the hadron contamination to the impact parameter distribution was absorbed into the Dalitz template by the fit method. To account for slight differences between the distributions, the result was compared with a fit using the hadron impact parameter template instead. A hypothetical template with the same mixture of Dalitz electrons and hadrons as in data would yield a result between these two extreme cases. For \pt~$\geq$~5~\GeVc, the fit using the hadron template was used for the central points as the contribution from hadrons dominates compared to that of the Dalitz electrons. The difference in the measured yield of electrons from beauty-hadron decays after exchanging the Dalitz template for the hadron template is 7\% at \pt~=~1.3~\GeVc decreasing towards higher transverse momentum. The proton contamination is significant only below \pt~=~1.3~\GeVc.

Like for the \pPb analysis, the influence of the difference in yield of mismatched conversions in data and Monte Carlo simulations had to be considered, especially as it increases with the multiplicity of the event. By making use of the multiplicity dependence, it was possible to create templates that either over- or underestimate this effect. This was cross-checked using charged pions from \KS decays as done in the \pPb analysis (see Section~\ref{pPbsys}). The change of the resulting measured spectra of electrons from beauty-hadron decays was used as an estimate for the systematic uncertainty, which is 14\% at \pt~=~1.3~\GeVc and decreases quickly towards higher transverse momentum.

As for the \pPb analysis, electrons from secondary pion and three-body decays of hadrons carrying a strange (or anti-strange) quark had to be considered, especially as these have broader impact parameter distributions than Dalitz electrons (see Section~\ref{Ana}). Due to the different final states, both the template for electrons from photon conversions and the template for Dalitz electrons are affected. These were split into a contribution from the decay of strange particles and the rest. For the fit they were considered as separate templates, but the amplitude parameters were coupled to have a fixed ratio. This was necessary because the contribution from strangeness is very small and could not be constrained by the information from the impact parameter distribution alone. The relative strength of the strangeness content was varied by a factor of two which includes the variation expected from the measured kaon/pion ratio~\cite{Abelev:2014laa}. The resulting difference in the yield of electrons from beauty-hadron decays was used as the estimate for the systematic uncertainty. It is $1.3\%$ for low \pt, decreasing towards higher transverse momentum.

Electrons at a fixed transverse momentum have mother particles in a range of \pt values. The impact parameter distributions of electrons depend on the momentum distributions of the mother particles. For the charm case this can be disentangled by making use of the measured charm \pt distribution~\cite{ALICE:2012ab}. For the beauty case this means that the result of the measurement depends on the input beauty-hadron 
spectrum in the Monte Carlo simulation. The effect was estimated by varying the beauty-hadron \pt distribution of the templates and observing the resulting change in the measured electron \pt distribution. The beauty-hadron \pt distribution was obtained according to PYTHIA simulations with a Perugia-0 tune which describes the measured \pPb data well. Therefore, an effect of the variation of the \pt distribution was studied 
by introducing a momentum-dependent nuclear modification factor \RAA. An \RAA based on a theoretical calculation was used for the central points~\cite{He:2014cla}. It has values near unity for low transverse momenta and drops to about $0.5$ from a hadron \pt of $5$ to $10~\rm{GeV}/c$. This was varied to half its effect ($\RAA \rightarrow (1+\RAA)/2$) in order to estimate the associated uncertainty. For the charm case, the variation was done according to the measurement uncertainties~\cite{ALICE:2012ab}. The difference in the resulting measured yield of electrons from beauty-hadron decays is about $8\%$, with no visible \pt dependence.

For the template fit, all species of charmed hadrons were combined into one template. The same holds for the beauty case. The baryon fraction of heavy hadrons is currently not known for \PbPb collisions and might be different than for pp collisions. Because of the different masses and decay channels, the various heavy-flavour hadron decays produce electrons with different impact parameter distributions. The templates were split into their contributions from only mesons or only baryons, with fixed ratios of the fit amplitudes. To estimate the uncertainty, the baryon fraction was increased by a factor of three for both charm and beauty simultaneously, motivated by the results of thermal model calculations~\cite{Oh:2009zj}. This led to a change in the measured yield of electrons from beauty-hadron decays of about $5\%$ with no clear momentum dependence. Decreasing the baryon ratio even to 0 has a smaller effect.

\begin{table}[h]
\centering
\scalebox{0.9}{
\begin{tabular}{|l|c|}

\hline
Source                           & Associated uncertainty\\
\hline
Tracking and matching            &  ~4.7\%  \\
TOF matching and eID             &  10\% \\ 
TPC eID                          &  ~5\%  \\
\hline
Signal extraction                & 17\% to 12\% \\
$d_0$ resolution correction       & 10\% to 0.4\%  \\
Hadron contamination             & 7\% to 1.4\%  \\
Mismatched conversions           & 14\% to 0.02\%  \\
Strangeness                      & ~1.3\% to 0.3\%   \\

Mother particle \pt distribution & ~8\% \\
Baryon/meson ratio               & ~5\% \\
\hline
\hline
Total                            & 26\% to 17\%  \\
\hline
\end{tabular}
}
\caption{Systematic uncertainties in the \PbPb analysis. Individual sources of systematic uncertainties are \pt dependent, which is reported using intervals. The lower and upper value of the interval, respectively, lists the uncertainty at \pt~=~1.3~\GeVc and \pt~=~8.0~\GeVc. The second group of entries in the table is related to the method used to extract the electrons from beauty-hadron decays.}
\label{tab::systPbPb}
\end{table}

\section{Reference pp cross sections at $\mathbf{\s}$~=~2.76~TeV and $\mathbf{\s}$~=~5.02~TeV}\label{ppref}

For the calculations of the nuclear modification factors \RpPb and \RPbPb, corresponding pp reference spectra at \s~=~5.02~TeV and \s~=~2.76~TeV are needed. To obtain these, the same method is used in both analyses. It is described in more detail in the following for the \pPb analysis. 

At present no pp measurement at \s~=~5.02~TeV exists. Therefore, the cross section of electrons from beauty-hadron decays measured in the momentum interval \ptrange{1}{8} at \s~=~7~TeV~\cite{Abelev:2012sca} was scaled to \s~=~5.02~TeV by applying a pQCD-driven \s-scaling~\cite{Averbeck:2011ga}. 
The \pt-dependent scaling function was obtained by calculating the ratio of the production cross sections of electrons from beauty-hadron decays from FONLL pQCD calculations~\cite{Cacciari:1998it,Cacciari:2001td,Cacciari:2012ny} at \s~=~5.02~TeV and \s~=~7~TeV. Both the direct ($\rm{b} \rightarrow \rm{e}$) and the cascade decay ($\rm{b} \rightarrow \rm{c} \rightarrow \rm{e}$) were considered. For the calculations at both energies the same parameters  were used for the beauty-quark mass ($m_{\rm b} = 4.75$~\GeVcc), the PDFs (CTEQ6.6~\cite{Nadolsky:2008zw}) as well as the factorisation $\mu_{\rm F}$ and renormalisation $\mu_{\rm R}$ scales with $\mu_{\rm R}$ = $\mu_{\rm F}$ = $\mu_{\rm 0}$ = $\sqrt{m_{\rm b}^2+p_{\rm T,b}^2}$, where $p_{\rm T,b}$ denotes the transverse momentum of the beauty quark. The uncertainties of the \pt-dependent scaling function were estimated by varying the parameters. The beauty-quark mass was set to $m_{\rm b} = $~4.5 and 5~\GeVcc. The uncertainties for the PDFs were obtained by using the CTEQ6.6 PDF uncertainties~\cite{Nadolsky:2008zw}.
The contribution from the scale uncertainties was estimated by using six different sets: {($\mu_{\rm R}/\mu_0,\mu_{\rm F}/\mu_0$)} = {(0.5,0.5),(1,0.5),(0.5,1),(2,1),(1,2),(2,2)}. The uncertainties originating from the mass and PDF variations are negligible. The uncertainty stemming from the variation of the scales was defined as the largest deviation from the scaling factor obtained with $\mu_{\rm R}$ = $\mu_{\rm F}$ = $\mu_{\rm 0}$. The resulting \s-scaling uncertainty is almost independent of \pt. It ranges from $^{+4}_{-2}$\% at 1~\GeVc to about $^{+2}_{-2}$\% at 8~\GeVc. The total systematic uncertainty of the pp reference spectrum at \s~=~5.02~TeV is then given as the bin-by-bin quadratic sum of the \s-scaling uncertainty and the relative systematic uncertainty of the measured spectrum at \s~=~7~TeV. For the statistical uncertainties the relative uncertainties of the spectrum measured at \s~=~7~TeV were taken. 

\begin{figure}[bt]
\centering
  \includegraphics[scale=0.42]{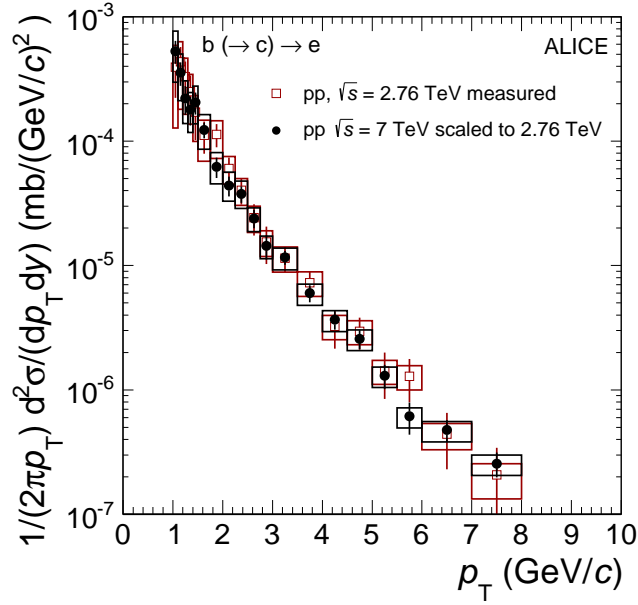}
  \caption{Invariant cross section of electrons from beauty-hadron decays at \s~=~2.76~TeV obtained by a pQCD-driven scaling of the cross section measured in pp collisions at \s~=~7~TeV in comparison with the measured spectrum in pp collisions at \s~=~2.76~TeV~\cite{Abelev:2014hla}. 
}
\label{fig::ppref}
\end{figure}

For the \RPbPb analysis, the measured spectrum at \s~=~7~TeV was scaled to \s~=~2.76~TeV using FONLL pQCD calculations at the respective energies. The systematic scaling uncertainty is about $^{+11}_{-\phantom{a}7}$\% at 1~\GeVc and about $^{+7}_{-5}$\% at 8~\GeVc. The resulting pp reference spectrum was found to be consistent with the measurement of electrons from beauty-hadron decays in pp collisions at \s~=~2.76~TeV~\cite{Abelev:2014hla}, shown in Fig.~\ref{fig::ppref}. The measured spectrum at \s~=~2.76~TeV was not taken as a reference for the \RPbPb, because of larger statistical and systematic uncertainties than the reference obtained via the \s-scaling. 

The systematic uncertainty of the normalisation related to the determination of the cross section of the minimum-bias trigger used for the measurement at \s~=~7~TeV is 3.5\% and also holds for the obtained pp reference spectra at \s~=~5.02~TeV and \s~=~2.76~TeV. 

The systematic uncertainties of the input \pt-differential cross section of electrons from beauty-hadron decays measured at \s~=~7~TeV, the normalisation uncertainty, as well as the scaling uncertainties for the reference spectra are summarised in Table~\ref{tab::systppref}.

\begin{table}[h]
\centering
\scalebox{0.9}{
\begin{tabular}{|l|c|c|c|}

\hline
pp spectrum 7 TeV  & 
\multicolumn{2}{c|}{\begin{tabular}[c]{@{}l@{}}45\% to 35\% for \ptrange{1}{1.5}\\35\% to 20\% for \ptrange{1.5}{2.5} \\ ~~~~~~~~~$\leq$~20\% for \pt~$\ge$~2.5~\GeVc \end{tabular}} \\
Normalisation uncertainty  &  
\multicolumn{2}{c|}{3.5\%}  \\
\hline
scaling uncertainty for & \pPb (\s~=~5.02~TeV)  & \PbPb (\s~=~2.76~TeV) \\
at \pt~=~1~\GeVc  &  $^{+4}_{-2}$\%   &   $^{+11}_{-\phantom{a}7}$\%  \\
at \pt~=~8~\GeVc  &  $^{+2}_{-2}$\%    &   ~$^{+7}_{-5}$\%        \\
\hline
\end{tabular}
}
\caption{Systematic uncertainties of the \pt-differential cross section of electrons from beauty-hadron decays measured at \s~=~7~TeV~\cite{Abelev:2012sca}, the normalisation uncertainty, as well as the scaling uncertainties for the reference spectra at \s~=~5.02~TeV and \s~=~2.76~TeV. The scaling uncertainties for the reference spectra are slightly \pt dependent; the uncertainties are given for the two extreme \pt intervals. Details are described in the text.}
\label{tab::systppref}
\end{table}

\begin{figure}[hbt]
\centering
 \includegraphics[width=0.48\textwidth]{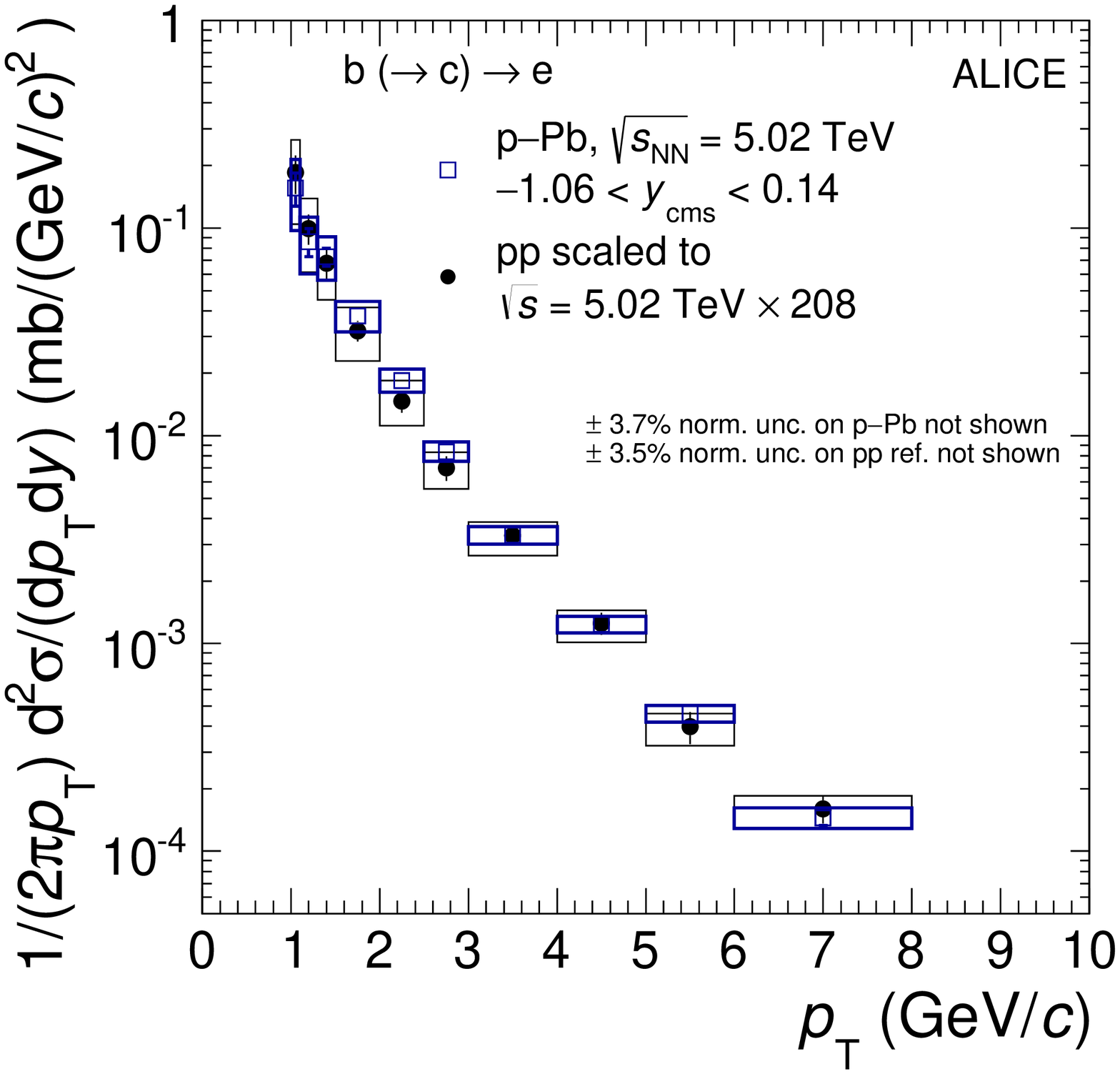}\hspace{0.5cm}
 \includegraphics[width=0.48\textwidth]{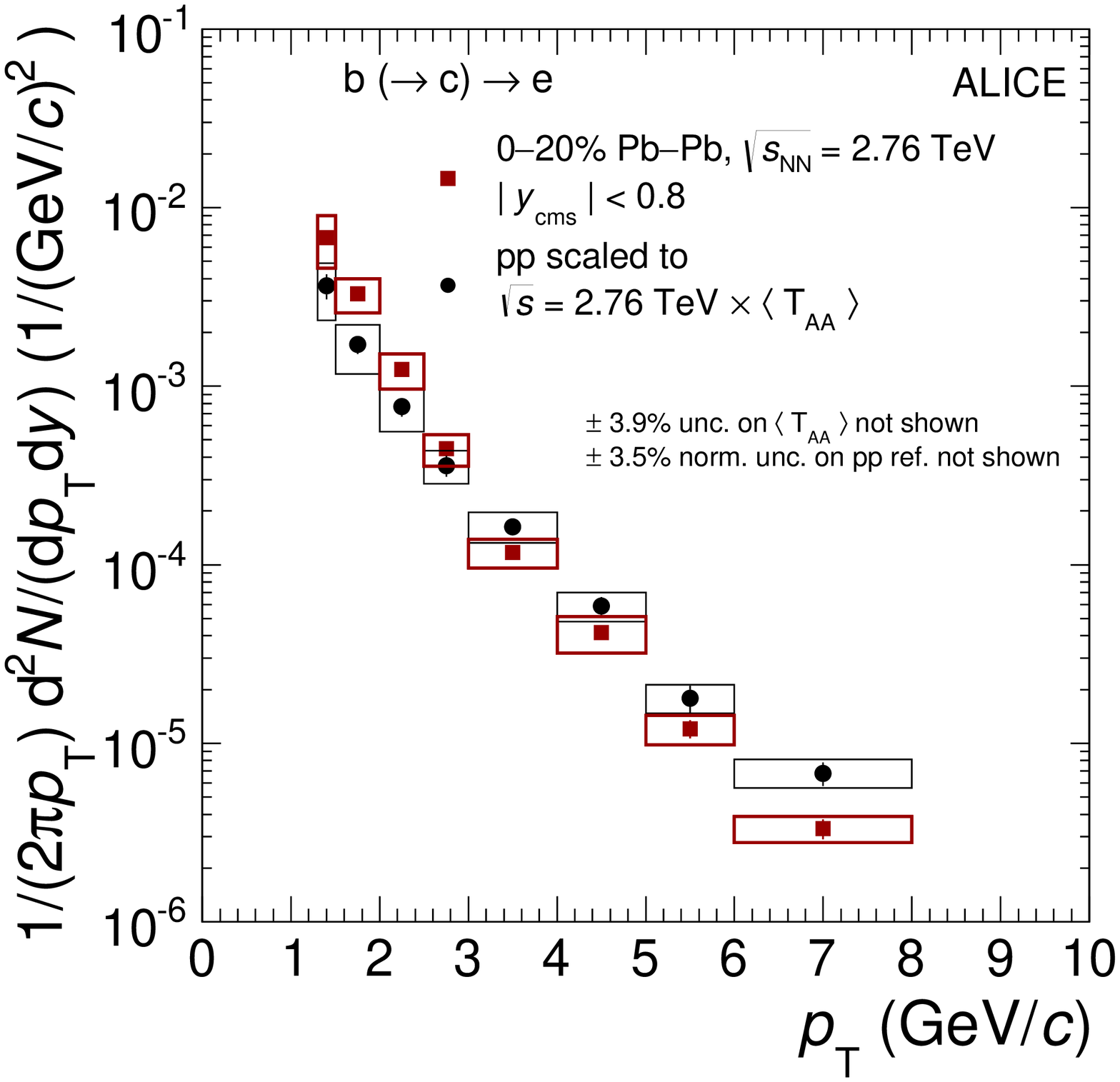}
  \caption{Invariant cross section (left) and yield (right) of electrons from beauty-hadron decays as a function of transverse momentum in minimum-bias \pPb collisions at \sqpA and in the 20\% most central \PbPb collisions at \sqAA. The pp reference spectra scaled by the number of nucleons in the Pb nucleus (A = 208) and by $\langle T_{\rm AA} \rangle$, respectively, are shown as well. The vertical bars represent the statistical uncertainties, the boxes indicate the systematic uncertainties. The pp and \pPb normalisation uncertainties of 3.5\% and 3.7\% as well as the one of the nuclear overlap function $\langle T_{\rm AA} \rangle$ of 3.9\% are not shown. }
\label{fig::hfept}
 \end{figure}

\section{Results}\label{Res}

The \pt-differential cross section and invariant yield of electrons from beauty-hadron decays at mid-rapidity in minimum-bias \pPb co\-llis\-ions at \sqpA and in the 20\% most central \PbPb collisions at \sqAA, respectively, are shown in Fig.~\ref{fig::hfept}. The markers are plotted at the centre of the \pt~bin. The vertical bars indicate the statistical uncertainties, the boxes represent the systematic uncertainties. The pp reference spectra, obtained via the pQCD-driven \s-scaling from the measurement in pp collisions at \s~=~7~TeV  as described in Section~\ref{ppref}, are shown for comparison. The pp reference spectra were multiplied by the number of nucleons in the Pb nucleus (A = 208) for the \pPb and with the nuclear overlap function ($\langle T_{\rm AA} \rangle$) for the \PbPb comparison.
The \PbPb result shows a suppression of electrons from beauty-hadron decays at high~\pt compared with the yield in pp collisions. Such a suppression is not seen in the comparison of the \pPb spectrum with the corresponding pp reference.

\begin{figure}[bt]
\centering
  \includegraphics[width=0.48\textwidth]{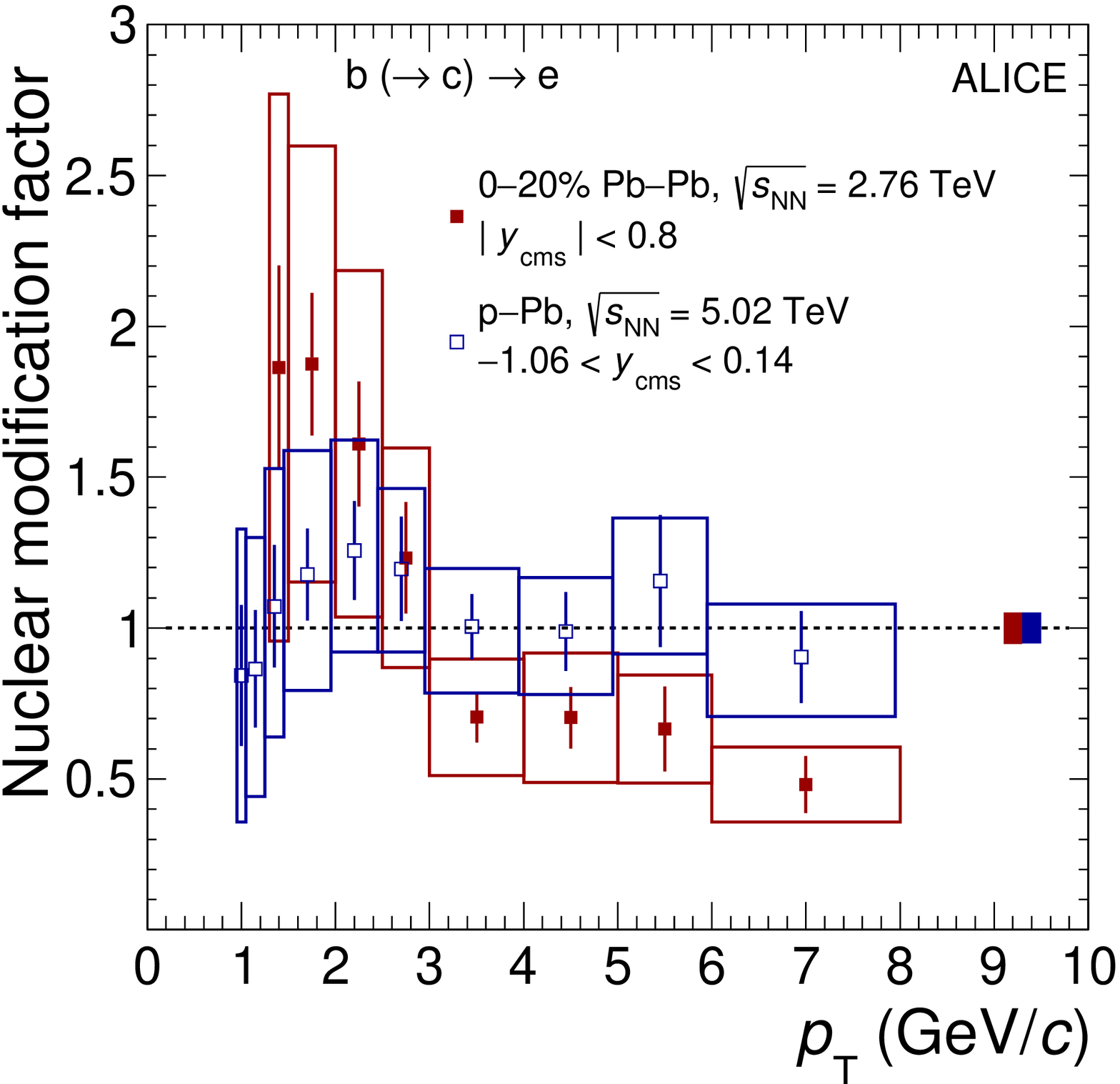}\hspace{0.5cm}
  \includegraphics[width=0.48\textwidth]{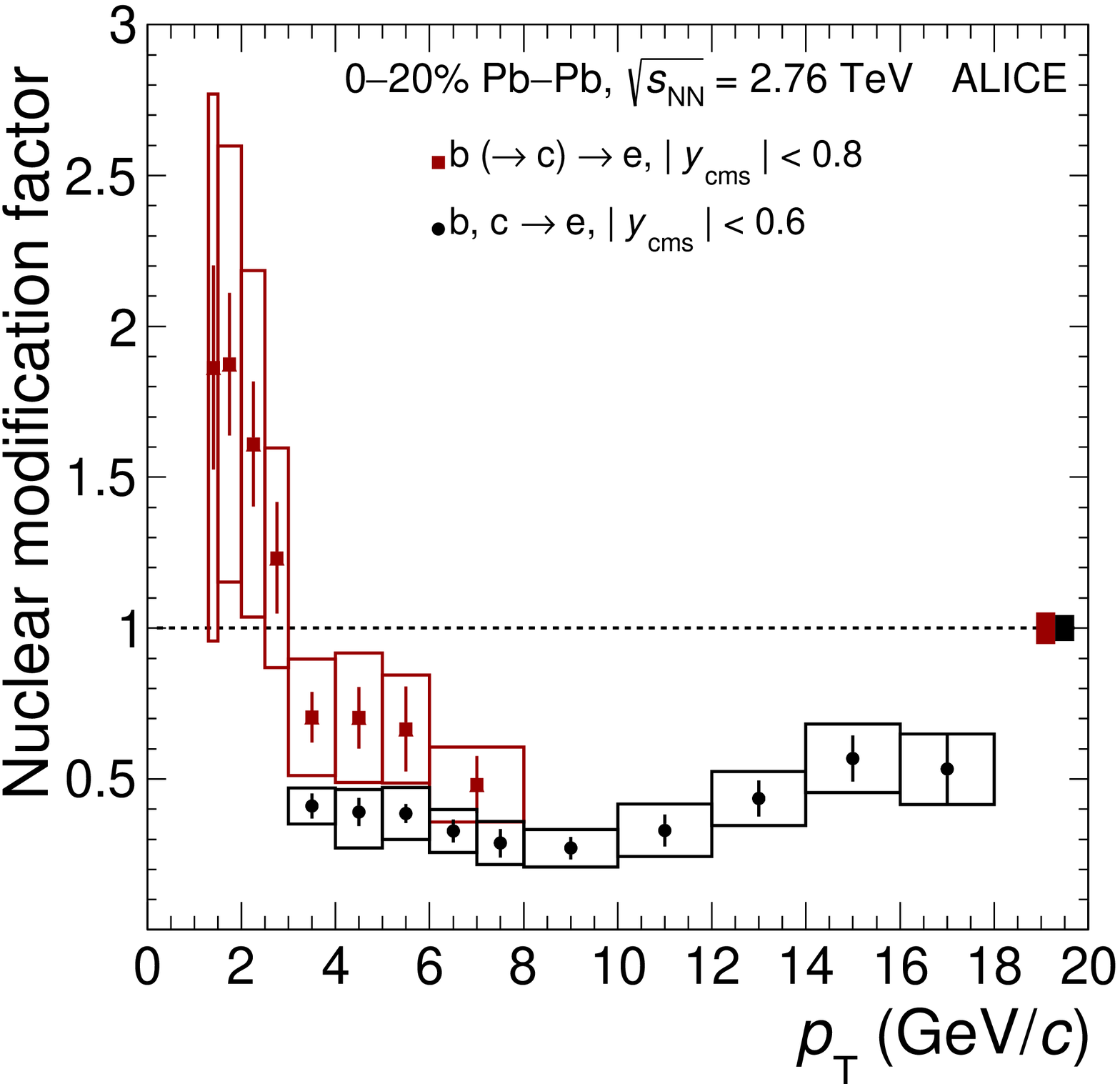}
  \caption{(left) Nuclear modification factors \RpPb and \RPbPb of electrons from beauty-hadron decays at mid-rapidity as a function of transverse momentum for minimum-bias \pPb collisions at \mbox{\sqpA} and 20\% most central \PbPb collisions at \mbox{\sqAA}. The data points of the \pPb analysis were shifted by 0.05~\GeVc to the left along the \pt axis for better visibility. (right) \RPbPb of electrons from beauty-hadron decays together with the corresponding result for beauty- and charm-hadron decays~\cite{InclusiveelectronRAA} for the 20\% most central \PbPb collisions. The vertical bars represent the statistical uncertainties, while the boxes indicate the systematic uncertainties. The normalisation uncertainties, common to all points, are shown as filled boxes at high \pt for all nuclear modification factors.  
  }
\label{fig::hferaa}
 \end{figure}

The nuclear modification factors \RPbPb and \RpPb are shown in Fig.~\ref{fig::hferaa} (left). The \RPbPb was obtained using Equation~\ref{RAAFormula}. The \RpPb was calculated as the ratio of the cross section of electrons from beauty-hadron decays in \pPb and pp collisions scaled by the number of nucleons in the Pb nucleus (A = 208). The statistical and systematic uncertainties of the \PbPb or \pPb and the pp spectra were propagated as independent uncertainties. The systematic uncertainties of the nuclear modification factors are partially correlated between the \pt bins. The normalisation uncertainty of the pp spectrum and the uncertainty of the nuclear overlap function $\langle T_{\rm AA} \rangle$ or the normalisation uncertainties of the \pPb spectrum, respectively, were added in quadrature. The normalisation uncertainties are shown as filled boxes at high transverse momentum in Fig.~\ref{fig::hferaa}.

The \RpPb is consistent with unity within uncertainties (of about 20\% for \pt~$>$ 2~\GeVc)  for all shown transverse momenta. The production of electrons from beauty-hadron decays is thus consistent with binary-collision scaling of the corresponding measurement in pp collisions at the same centre-of-mass energy. The values of the \RPbPb for the 20\% most central \PbPb collisions increase, for \pt~$\le$ 3~\GeVc, with sizeable uncertainties of 30--45\%.
In the interval \ptrange{3}{6}, the \RPbPb is about 0.7 with a systematic uncertainty of about 30\%; in \ptrange{6}{8} the ratio is 0.48 with an uncertainty of about 25\%. In the latter transverse momentum range the suppression with respect to \RPbPb~=~1 is a 3.3$\sigma$ effect taking into account the statistical and systematic uncertainties.

A comparison of the \RPbPb of electrons from beauty-hadron decays with the one from charm- and beauty-hadron decays is shown in Fig.~\ref{fig::hferaa} (right) for the 20\% most central \PbPb collisions. For the latter \RPbPb, the \pt-differential invariant yields of electrons from charm- and beauty-hadron decays published in~\cite{InclusiveelectronRAA} for the centrality classes 0--10\% and 10--20\% were combined. For the pp reference in the momentum range up to \pt~$\le$~12~\GeVc, the corresponding invariant cross section measurement at \s~=~2.76~TeV~\cite{Abelev:2014gla}, which has uncertainties of about 20\%, was used. For \pt~$\ge$~12~\GeVc, the ATLAS measurement~\cite{Abelev:2012xe} at \s~=~7~TeV was extrapolated to \s~=~2.76~TeV applying a FONLL pQCD-driven \s-scaling analogous to the method described in Section~\ref{ppref}. The uncertainty of the pp reference in this momentum range is about 15\%. As expected, the results agree within uncertainties at high \pt, where the beauty contribution is larger than the charm contribution~\cite{Abelev:2014gla}. In the \pt interval \ptrange{3}{6}, the suppression of the \RPbPb for electrons from beauty-hadron decays is about 1.2$\sigma$ less. This difference is consistent with the ordering of charm and beauty suppression seen in the prompt D meson and \jpsi from B meson comparison~\cite{Adam:2015nna,Adam:2015rba,Chatrchyan:2012np}.

\begin{figure}[bt]
\centering
 \includegraphics[width=0.48\textwidth]{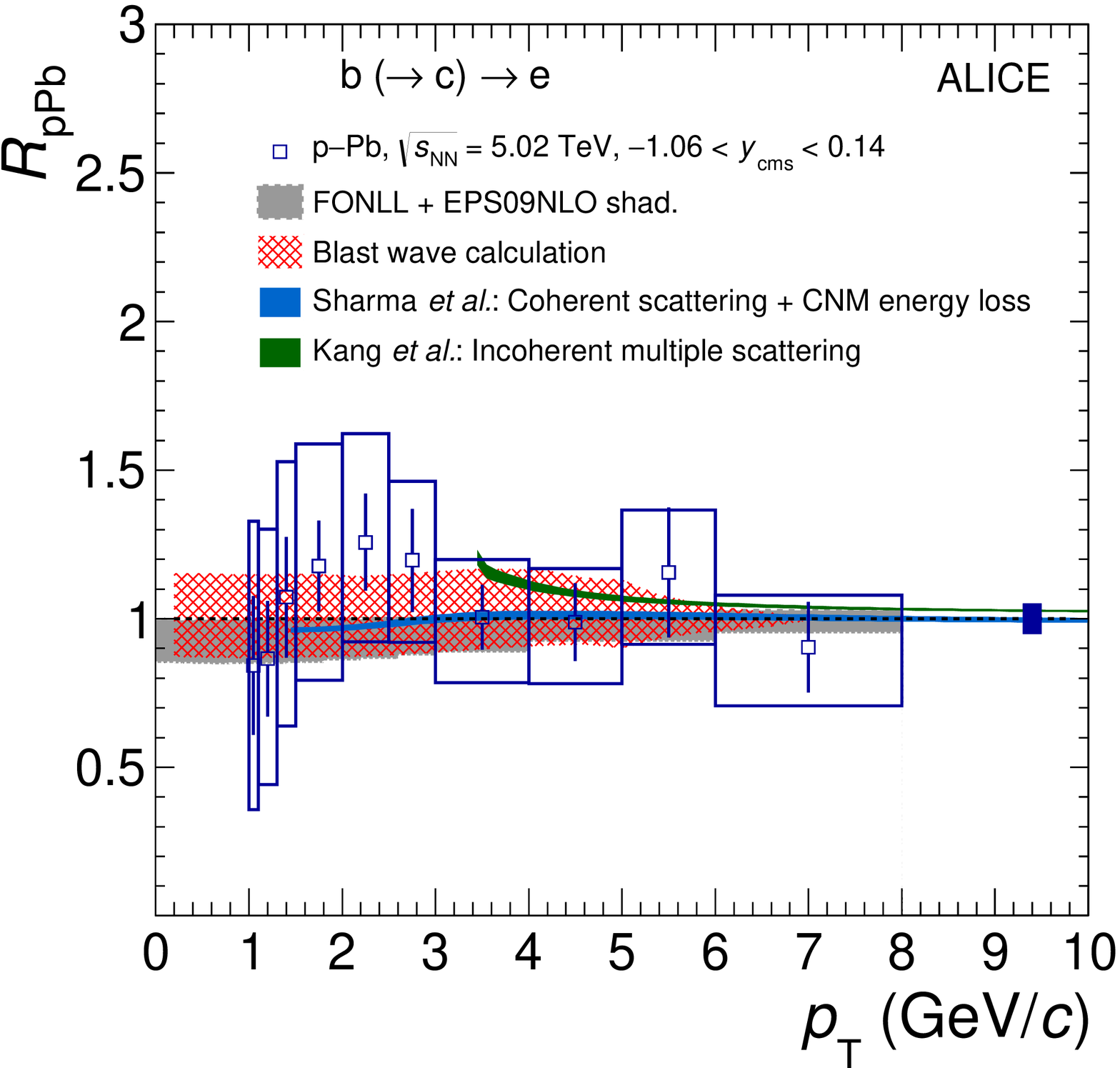}\hspace{0.5cm}
 \includegraphics[width=0.48\textwidth]{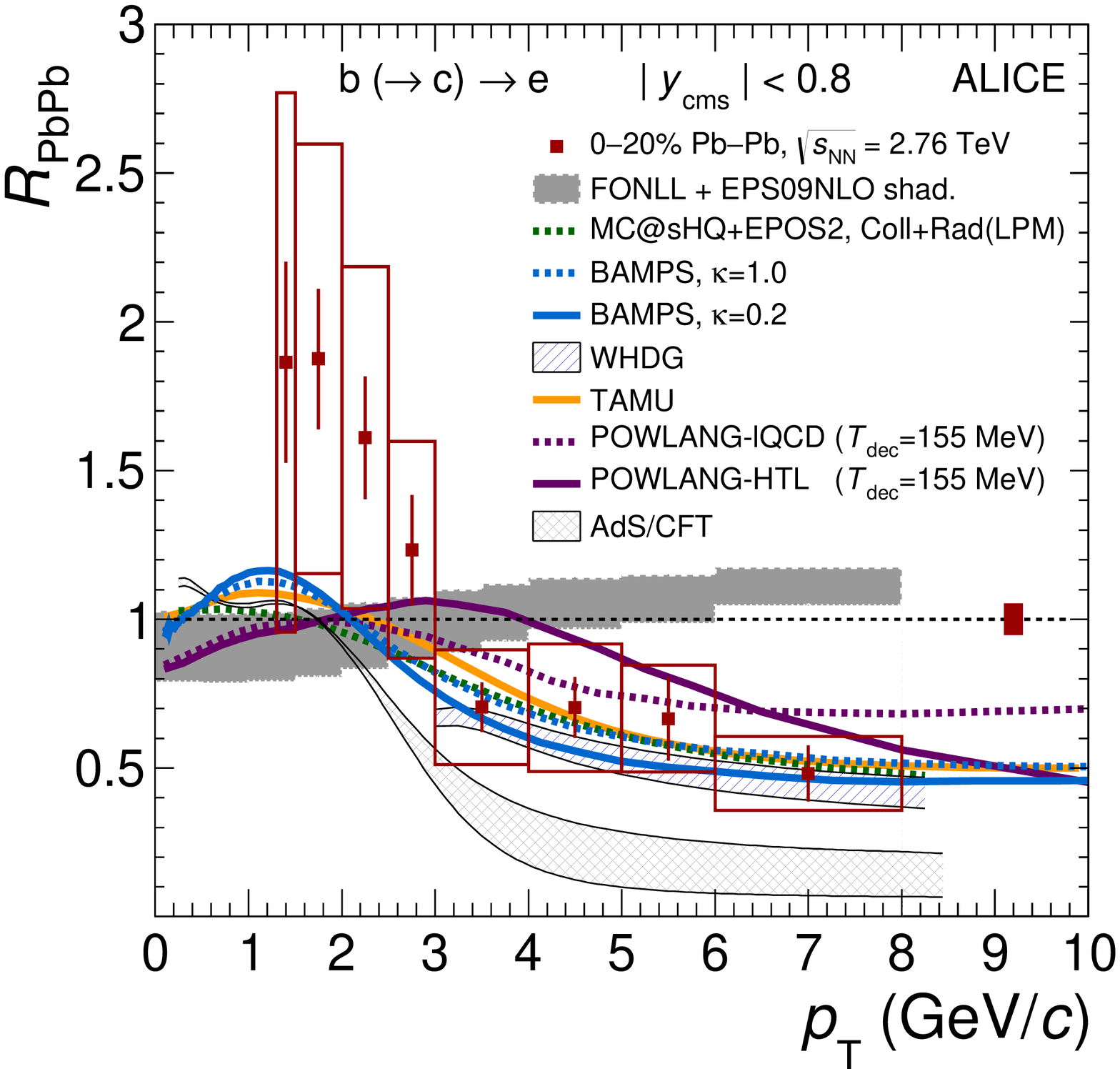}
 \caption{Nuclear modification factors \RpPb (left) and \RPbPb (right) of electrons from beauty-hadron decays in comparison with different theoretical predictions~\cite{Cacciari:1998it,Cacciari:2001td,Cacciari:2012ny,Eskola:2009uj,Sickles:2013yna,
Sharma:2009hn,Kang:2014hha,Nahrgang:2013xaa,He:2014cla,Beraudo:2014boa,Horowitz:2011gd,Wicks:2005gt,
Horowitz:2015dta,Uphoff:2014hza,Uphoff:2014cba}, see text for details. The vertical bars represent the statistical uncertainties, while the boxes indicate the systematic uncertainties. The normalisation uncertainty, common to all points, is shown as a filled box at high \pt for both collision systems. }\label{fig::hferaa_wtheory}
\end{figure}

Within uncertainties, the \RpPb is described by pQCD calculations including modifications of the parton distribution functions (FONLL~\cite{Cacciari:1998it,Cacciari:2001td,Cacciari:2012ny} + EPS09NLO~\cite{Eskola:2009uj} nuclear PDFs) as shown in Fig.~\ref{fig::hferaa_wtheory} (left). The data and the calculation suggest that cold nuclear matter effects are small at high transverse momentum. Recent measurements of long-range correlations for charged hadrons~\cite{Abelev:2012ola,Aad:2013fja,CMS:2012qk} and studies of the mean transverse momentum as a function of the charged-particle multiplicity in the event~\cite{Abelev:2013haa} suggest that there might be collective effects in \pPb collisions. The figure also reports the result of a calculation based on the idea proposed in Ref.~\cite{Sickles:2013yna}, in which the \pt distribution of beauty hadrons from a hydrodynamically expanding medium is obtained from a blast-wave model. The blast-wave parameters were extracted from fits to the \pt-spectra of light hadrons~\cite{Abelev:2013haa} in \pPb collisions. The uncertainties of the measurement do not allow for a conclusion on possible flow effects. The data are also described by calculations which include CNM energy loss, nuclear shadowing and coherent multiple scattering at the partonic level~\cite{Sharma:2009hn}. An enhancement at intermediate \pt is predicted by the calculations based on incoherent multiple scattering~\cite{Kang:2014hha}. Presently, the large systematic uncertainties of the measurement do not allow one to discriminate between the aforementioned theoretical approaches.

Perturbative QCD calculations including initial-state effects for \PbPb collisions at \s~=~2.76~TeV (FONLL~\cite{Cacciari:1998it,Cacciari:2001td,Cacciari:2012ny} + EPS09NLO~\cite{Eskola:2009uj} nuclear PDFs) cannot describe the \RPbPb at high transverse momentum (see Fig.~\ref{fig::hferaa_wtheory}, right), indicating that the suppression, particularly evident in the interval \ptrange{6}{8}, is induced by the presence of a hot and dense medium in the final state. At lower transverse momentum, the large uncertainties do not allow one to conclude whether the measured \RPbPb is larger than that obtained from this calculation. 

In order to gain further insight into the energy loss mechanisms, particularly the relative importance of radiative and collisional energy loss, the data are compared with several models of heavy-quark transport and energy loss in the QGP. Both radiative and collisional energy loss are included in the pQCD model MC@sHQ+EPOS2~\cite{Nahrgang:2013xaa}, the partonic transport description BAMPS~\cite{Uphoff:2014hza,Uphoff:2014cba}, and in WHDG~\cite{Horowitz:2011gd,Wicks:2005gt,Horowitz:2015dta}. The non-perturbative transport model TAMU~\cite{He:2014cla} includes only collisional processes, while the POWLANG~\cite{Beraudo:2014boa} transport calculation simulates the production of heavy quarks using POWHEG and their propagation in the plasma via a relativistic Langevin equation. Heavy-quark energy loss can also be calculated using the AdS/CFT heavy-quark drag
model~\cite{Horowitz:2015dta}.

The right-hand side of Fig.~\ref{fig::hferaa_wtheory} shows the comparison of the various models with the measured \RPbPb. The MC@sHQ+EPOS2 calculation with EPOS initial conditions~\cite{Werner:2010aa,Werner:2012xh}, including the Landau-Pomeranchuk-Migdal (LPM) effect~\cite{Baier:1994bd}, is consistent with the data at high~\pt. The BAMPS~\cite{Uphoff:2014hza,Uphoff:2014cba} model is based on pQCD cross sections including the running of the coupling and scaled by a constant factor $\kappa$. The two shown values of $\kappa$ cannot be distinguished given the uncertainties in the data. In the WHDG calculation, the medium density is assumed to be proportional to the charged particle multiplicity and a 1-D Bjorken-expansion is included. The WHDG model describes the measurement well within the restricted \pt~range shown.

The TAMU model includes collisional processes and incorporates resonance formation close to the critical temperature as well as diffusion of heavy-flavour mesons in the hadronic phase. The hydrodynamic expansion is constrained by \pt spectra and elliptic flow measurements of light hadrons. The calculations are consistent with the data at high~\pt, indicating a limited sensitivity of the current data to radiative energy loss effects. 
The POWLANG~\cite{Beraudo:2014boa} transport calculation takes into account initial-state nuclear effects via EPS09 modifications of the PDFs and describes the medium using an underlying hydrodynamical model. The transport coefficients used for the evolution of the heavy quark in the medium are either extracted from lattice-QCD calculations or Hard-Thermal-Loop (HTL) resummation~\cite{Braaten:1989mz} of medium effects. The hadronisation via in-vacuum fragmentation functions or via in-medium string-fragmentation routines occurs once the decoupling temperature is reached. The calculations are shown for different transport coefficients with a decoupling temperature $T_{\rm dec}$~=~155~MeV; the results with a temperature of $T_{\rm dec}$~=~170~MeV look similar. No scenario is clearly favoured by the current data set. The AdS/CFT model, which includes energy loss fluctuations in a realistic strong-coupling energy loss mode, clearly shows a stronger suppression than the measured \RPbPb.
 
The MC@sHQ+EPOS2, the BAMPS as well as the TAMU calculation describe the suppression seen in data at high transverse momentum. They also show an increase towards lower momentum reaching \RPbPb values around unity or slightly above. The data show a larger increase with decreasing transverse momentum, however exhibit large systematic and statistical uncertainties.

\section{Summary}\label{Summarytex}

The \pt-differential cross section and invariant yield of electrons from beauty-hadron decays in minimum-bias \pPb collisions and in the 20\% most central \PbPb collisions, respectively, were measured at mid-rapidity. The measurements are compared via the nuclear modification factors with pp reference spectra, obtained by a pQCD-driven \s-scaling of the cross section of electrons from beauty-hadron decays measured at \s~=~7~TeV. The \RpPb is consistent with unity within uncertainties of about 20\% at high transverse momentum \pt, which increase towards low \pt. The \RpPb is described by pQCD calculations including initial-state effects, energy loss approaches as well as by a blast wave model calculation that parametrises possible hydrodynamic effects. The \RPbPb is about 0.7 with an uncertainty of about 30\% in the interval \ptrange{3}{6} and 0.48 with an uncertainty of about 25\% for \ptrange{6}{8}. The suppression seen in the higher transverse momentum interval is not described by pQCD calculations including only initial-state effects, indicating a final-state effect as the origin. The values of the \RPbPb increase for \pt~$\le 3$~\GeVc with uncertainties of about 30--45\%. The measured \RPbPb is described within uncertainties by pQCD-inspired models of beauty-quark energy loss in the QGP. In the interval \ptrange{3}{6}, we observe that the suppression of the \RPbPb for electrons from beauty-hadron decays is about 1.2$\sigma$ less than that from charm- and beauty hadron decays. This difference is consistent with the ordering of charm and beauty suppression seen in the prompt D meson and \jpsi from B meson comparison.

%
%

\newenvironment{acknowledgement}{\relax}{\relax}
\begin{acknowledgement}
\section*{Acknowledgements}

The ALICE Collaboration would like to thank all its engineers and technicians for their invaluable contributions to the construction of the experiment and the CERN accelerator teams for the outstanding performance of the LHC complex.
The ALICE Collaboration gratefully acknowledges the resources and support provided by all Grid centres and the Worldwide LHC Computing Grid (WLCG) collaboration.
The ALICE Collaboration acknowledges the following funding agencies for their support in building and running the ALICE detector:
A. I. Alikhanyan National Science Laboratory (Yerevan Physics Institute) Foundation (ANSL), State Committee of Science and World Federation of Scientists (WFS), Armenia;
Austrian Academy of Sciences and Nationalstiftung f\"{u}r Forschung, Technologie und Entwicklung, Austria;
Conselho Nacional de Desenvolvimento Cient\'{\i}fico e Tecnol\'{o}gico (CNPq), Financiadora de Estudos e Projetos (Finep) and Funda\c{c}\~{a}o de Amparo \`{a} Pesquisa do Estado de S\~{a}o Paulo (FAPESP), Brazil;
Ministry of Education of China (MOE of China), Ministry of Science \& Technology of China (MOST of China) and National Natural Science Foundation of China (NSFC), China;
Ministry of Science, Education and Sport and Croatian Science Foundation, Croatia;
Centro de Investigaciones Energ\'{e}ticas, Medioambientales y Tecnol\'{o}gicas (CIEMAT), Cuba;
Ministry of Education, Youth and Sports of the Czech Republic, Czech Republic;
Danish National Research Foundation (DNRF), The Carlsberg Foundation and The Danish Council for Independent Research | Natural Sciences, Denmark;
Helsinki Institute of Physics (HIP), Finland;
Commissariat \`{a} l'Energie Atomique (CEA) and Institut National de Physique Nucl\'{e}aire et de Physique des Particules (IN2P3) and Centre National de la Recherche Scientifique (CNRS), France;
Bundesministerium f\"{u}r Bildung, Wissenschaft, Forschung und Technologie (BMBF) and GSI Helmholtzzentrum f\"{u}r Schwerionenforschung GmbH, Germany;
Ministry of Education, Research and Religious Affairs, Greece;
National Research, Development and Innovation Office, Hungary;
Department of Atomic Energy Government of India (DAE), India;
Indonesian Institute of Science, Indonesia;
Centro Fermi - Museo Storico della Fisica e Centro Studi e Ricerche Enrico Fermi and Istituto Nazionale di Fisica Nucleare (INFN), Italy;
Institute for Innovative Science and Technology , Nagasaki Institute of Applied Science (IIST), Japan Society for the Promotion of Science (JSPS) KAKENHI and Japanese Ministry of Education, Culture, Sports, Science and Technology (MEXT), Japan;
Consejo Nacional de Ciencia (CONACYT) y Tecnolog\'{i}a, through Fondo de Cooperaci\'{o}n Internacional en Ciencia y Tecnolog\'{i}a (FONCICYT) and Direcci\'{o}n General de Asuntos del Personal Academico (DGAPA), Mexico;
Nationaal instituut voor subatomaire fysica (Nikhef), Netherlands;
The Research Council of Norway, Norway;
Commission on Science and Technology for Sustainable Development in the South (COMSATS), Pakistan;
Pontificia Universidad Cat\'{o}lica del Per\'{u}, Peru;
Ministry of Science and Higher Education and National Science Centre, Poland;
Ministry of Education and Scientific Research, Institute of Atomic Physics and Romanian National Agency for Science, Technology and Innovation, Romania;
Joint Institute for Nuclear Research (JINR), Ministry of Education and Science of the Russian Federation and National Research Centre Kurchatov Institute, Russia;
Ministry of Education, Science, Research and Sport of the Slovak Republic, Slovakia;
National Research Foundation of South Africa, South Africa;
Korea Institute of Science and Technology Information and National Research Foundation of Korea (NRF), South Korea;
Centro de Investigaciones Energ\'{e}ticas, Medioambientales y Tecnol\'{o}gicas (CIEMAT) and Ministerio de Ciencia e Innovacion, Spain;
Knut \& Alice Wallenberg Foundation (KAW) and Swedish Research Council (VR), Sweden;
European Organization for Nuclear Research, Switzerland;
National Science and Technology Development Agency (NSDTA), Office of the Higher Education Commission under NRU project of Thailand and Suranaree University of Technology (SUT), Thailand;
Turkish Atomic Energy Agency (TAEK), Turkey;
National Academy of  Sciences of Ukraine, Ukraine;
Science and Technology Facilities Council (STFC), United Kingdom;
National Science Foundation of the United States of America (NSF) and United States Department of Energy, Office of Nuclear Physics (DOE NP), United States.    
\end{acknowledgement}

\clearpage
\bibliographystyle{utphys}
\bibliography{hfe_pPb_PbPb_btoe.bib}

\providecommand{\href}[2]{#2}\begingroup\raggedright\begin{thebibliography}{100}

\bibitem{Borsanyi:2010bp}
{\bfseries Wuppertal-Budapest} Collaboration, S.~Borsanyi, Z.~Fodor,
  C.~Hoelbling, S.~D. Katz, S.~Krieg, C.~Ratti, and K.~K. Szabo, ``{Is there
  still any $\rm T_c$ mystery in lattice QCD? Results with physical masses in
  the continuum limit III},''
  \href{http://dx.doi.org/10.1007/JHEP09(2010)073}{{\em JHEP} {\bfseries 09}
  (2010) 073},
\href{http://arxiv.org/abs/1005.3508}{{\ttfamily arXiv:1005.3508 [hep-lat]}}.

\bibitem{Bazavov:2014pvz}
{\bfseries HotQCD} Collaboration, A.~Bazavov {\em et~al.}, ``{Equation of state
  in ( 2+1 )-flavor QCD},''
  \href{http://dx.doi.org/10.1103/PhysRevD.90.094503}{{\em Phys. Rev.}
  {\bfseries D90} (2014) 094503},
\href{http://arxiv.org/abs/1407.6387}{{\ttfamily arXiv:1407.6387 [hep-lat]}}.

\bibitem{Anton:2015}
A.~Andronic {\em et~al.}, ``{Heavy-flavour and quarkonium production in the LHC
  era: from proton\mbox{-}proton to heavy-ion collisions},''
  \href{http://dx.doi.org/10.1140/epjc/s10052-015-3819-5}{{\em Eur. Phys. J.}
  {\bfseries C76} (2016) 107},
\href{http://arxiv.org/abs/1506.03981}{{\ttfamily arXiv:1506.03981 [nucl-ex]}}.

\bibitem{Liu:2012ax}
F.-M. Liu and S.-X. Liu, ``{Quark-gluon plasma formation time and direct
  photons from heavy ion collisions},''
  \href{http://dx.doi.org/10.1103/PhysRevC.89.034906}{{\em Phys. Rev.}
  {\bfseries C89} (2014) 034906},
\href{http://arxiv.org/abs/1212.6587}{{\ttfamily arXiv:1212.6587 [nucl-th]}}.

\bibitem{Aamodt:2011mr}
{\bfseries ALICE} Collaboration, K.~Aamodt {\em et~al.}, ``{Two-pion
  Bose-Einstein correlations in central Pb--Pb collisions at $\sqrt{s_{NN}} =$
  2.76 TeV},'' \href{http://dx.doi.org/10.1016/j.physletb.2010.12.053}{{\em
  Phys. Lett.} {\bfseries B696} (2013) 328--337},
\href{http://arxiv.org/abs/1012.4035}{{\ttfamily arXiv:1012.4035 [nucl-ex]}}.

\bibitem{Gyulassy:1993hr}
M.~Gyulassy and X.-N. Wang, ``{Multiple collisions and induced gluon
  bremsstrahlung in QCD},''
  \href{http://dx.doi.org/10.1016/0550-3213(94)90079-5}{{\em Nucl. Phys.}
  {\bfseries B420} (1994) 583--614},
\href{http://arxiv.org/abs/nucl-th/9306003}{{\ttfamily arXiv:nucl-th/9306003
  [nucl-th]}}.

\bibitem{Baier:1996sk}
R.~Baier, Y.~L. Dokshitzer, A.~H. Mueller, S.~Peigne, and D.~Schiff,
  ``{Radiative energy loss and p(T) broadening of high-energy partons in
  nuclei},'' \href{http://dx.doi.org/10.1016/S0550-3213(96)00581-0}{{\em Nucl.
  Phys.} {\bfseries B484} (1997) 265--282},
\href{http://arxiv.org/abs/hep-ph/9608322}{{\ttfamily arXiv:hep-ph/9608322
  [hep-ph]}}.

\bibitem{Wiedemann:2000za}
U.~A. Wiedemann, ``{Gluon radiation off hard quarks in a nuclear environment:
  opacity expansion},''
  \href{http://dx.doi.org/10.1016/S0550-3213(00)00457-0}{{\em Nucl. Phys.}
  {\bfseries B588} (2000) 303--344},
\href{http://arxiv.org/abs/hep-ph/0005129}{{\ttfamily arXiv:hep-ph/0005129
  [hep-ph]}}.

\bibitem{Dokshitzer:2001zm}
Y.~L. Dokshitzer and D.~E. Kharzeev, ``{Heavy quark colorimetry of QCD
  matter},'' \href{http://dx.doi.org/10.1016/S0370-2693(01)01130-3}{{\em Phys.
  Lett.} {\bfseries B519} (2001) 199--206},
\href{http://arxiv.org/abs/hep-ph/0106202}{{\ttfamily arXiv:hep-ph/0106202
  [hep-ph]}}.

\bibitem{Armesto:2003jh}
N.~Armesto, C.~A. Salgado, and U.~A. Wiedemann, ``{Medium induced gluon
  radiation off massive quarks fills the dead cone},''
  \href{http://dx.doi.org/10.1103/PhysRevD.69.114003}{{\em Phys. Rev.}
  {\bfseries D69} (2004) 114003},
\href{http://arxiv.org/abs/hep-ph/0312106}{{\ttfamily arXiv:hep-ph/0312106
  [hep-ph]}}.

\bibitem{PhysRevLett.93.072301}
B.-W. Zhang, E.~Wang, and X.-N. Wang, ``{Heavy Quark Energy Loss in Nuclear
  Medium},'' \href{http://dx.doi.org/10.1103/PhysRevLett.93.072301}{{\em Phys.
  Rev. Lett.} {\bfseries 93} (2004) 072301},
\href{http://arxiv.org/abs/nucl-th/0309040}{{\ttfamily arXiv:nucl-th/0309040
  [nucl-th]}}.

\bibitem{Djordjevic:2003zk}
M.~Djordjevic and M.~Gyulassy, ``{Heavy quark radiative energy loss in QCD
  matter},'' \href{http://dx.doi.org/10.1016/j.nuclphysa.2003.12.020}{{\em
  Nucl. Phys.} {\bfseries A733} (2004) 265--298},
\href{http://arxiv.org/abs/nucl-th/0310076}{{\ttfamily arXiv:nucl-th/0310076
  [nucl-th]}}.

\bibitem{Armesto:2005iq}
N.~Armesto, A.~Dainese, C.~A. Salgado, and U.~A. Wiedemann, ``{Testing the
  color charge and mass dependence of parton energy loss with heavy-to-light
  ratios at RHIC and CERN LHC},''
  \href{http://dx.doi.org/10.1103/PhysRevD.71.054027}{{\em Phys. Rev.}
  {\bfseries D71} (2005) 054027},
\href{http://arxiv.org/abs/hep-ph/0501225}{{\ttfamily arXiv:hep-ph/0501225
  [hep-ph]}}.

\bibitem{Armesto:2006ph}
N.~Armesto, ``{Nuclear shadowing},''
  \href{http://dx.doi.org/10.1088/0954-3899/32/11/R01}{{\em J. Phys.}
  {\bfseries G32} (2006) R367--R394},
\href{http://arxiv.org/abs/hep-ph/0604108}{{\ttfamily arXiv:hep-ph/0604108
  [hep-ph]}}.

\bibitem{Fujii:2006ab}
H.~Fujii, F.~Gelis, and R.~Venugopalan, ``{Quark pair production in high energy
  pA collisions: General features},''
  \href{http://dx.doi.org/10.1016/j.nuclphysa.2006.09.012}{{\em Nucl. Phys.}
  {\bfseries A780} (2006) 146--174},
\href{http://arxiv.org/abs/hep-ph/0603099}{{\ttfamily arXiv:hep-ph/0603099
  [hep-ph]}}.

\bibitem{Cronin:1974zm}
J.~W. Cronin, H.~J. Frisch, M.~J. Shochet, J.~P. Boymond, R.~Mermod, P.~A.
  Piroue, and R.~L. Sumner, ``{Production of hadrons with large transverse
  momentum at 200, 300, and 400 GeV},''
\href{http://dx.doi.org/10.1103/PhysRevD.11.3105}{{\em Phys. Rev.} {\bfseries
  D11} (1975) 3105}.

\bibitem{Sharma:2009hn}
R.~Sharma, I.~Vitev, and B.-W. Zhang, ``{Light-cone wave function approach to
  open heavy flavor dynamics in QCD matter},''
  \href{http://dx.doi.org/10.1103/PhysRevC.80.054902}{{\em Phys. Rev.}
  {\bfseries C80} (2009) 054902},
\href{http://arxiv.org/abs/0904.0032}{{\ttfamily arXiv:0904.0032 [hep-ph]}}.

\bibitem{Kang:2014hha}
Z.-B. Kang, I.~Vitev, E.~Wang, H.~Xing, and C.~Zhang, ``{Multiple scattering
  effects on heavy meson production in p+A collisions at backward rapidity},''
  \href{http://dx.doi.org/10.1016/j.physletb.2014.11.024}{{\em Phys. Lett.}
  {\bfseries B740} (2015) 23--29},
\href{http://arxiv.org/abs/1409.2494}{{\ttfamily arXiv:1409.2494 [hep-ph]}}.

\bibitem{Miller:2007ri}
M.~L. Miller, K.~Reygers, S.~J. Sanders, and P.~Steinberg, ``{Glauber Modeling
  in High Energy Nuclear Collisions},''
  \href{http://dx.doi.org/10.1146/annurev.nucl.57.090506.123020}{{\em Ann. Rev.
  Nucl. Part. Sci.} {\bfseries 57} (2007) 205--243},
\href{http://arxiv.org/abs/nucl-ex/0701025}{{\ttfamily arXiv:nucl-ex/0701025
  [nucl-ex]}}.

\bibitem{Aggarwal:2010xp}
{\bfseries STAR} Collaboration, M.~M. Aggarwal {\em et~al.}, ``{Measurement of
  the Bottom Quark Contribution to Nonphotonic Electron Production in $p+p$
  Collisions at $\sqrt{s} $=200 GeV},''
  \href{http://dx.doi.org/10.1103/PhysRevLett.105.202301}{{\em Phys. Rev.
  Lett.} {\bfseries 105} (2010) 202301},
\href{http://arxiv.org/abs/1007.1200}{{\ttfamily arXiv:1007.1200 [nucl-ex]}}.

\bibitem{Agakishiev:2011mr}
{\bfseries STAR} Collaboration, H.~Agakishiev {\em et~al.}, ``{High $p_{T}$
  non-photonic electron production in $p+p$ collisions at $\sqrt{s} = 200$
  GeV},'' \href{http://dx.doi.org/10.1103/PhysRevD.83.052006}{{\em Phys. Rev.}
  {\bfseries D83} (2011) 052006},
\href{http://arxiv.org/abs/1102.2611}{{\ttfamily arXiv:1102.2611 [nucl-ex]}}.

\bibitem{PhysRevD.71.032001}
{\bfseries CDF} Collaboration, D.~Acosta {\em et~al.}, ``{Measurement of the
  $J/\ensuremath{\psi}$ meson and $b$-hadron production cross sections in
  $p\bar{p}$ collisions at $\sqrt{s} = 1.96$ TeV},''
  \href{http://dx.doi.org/10.1103/PhysRevD.71.032001}{{\em Phys. Rev.}
  {\bfseries D71} (2005) 032001},
  \href{http://arxiv.org/abs/0412071}{{\ttfamily arXiv:0412071 [hep-ex]}}.

\bibitem{Abelev:2012sca}
{\bfseries ALICE} Collaboration, B.~Abelev {\em et~al.}, ``{Measurement of
  electrons from beauty hadron decays in $pp$ collisions at $\sqrt{s}=7$
  TeV},'' \href{http://dx.doi.org/10.1016/j.physletb.2016.10.004,
  10.1016/j.physletb.2013.01.069}{{\em Phys. Lett.} {\bfseries B721} (2013)
  13--23}, \href{http://arxiv.org/abs/1208.1902}{{\ttfamily arXiv:1208.1902
  [hep-ex]}}.
[Erratum: Phys. Lett.B763,507(2016)].

\bibitem{Abelev:2014gla}
{\bfseries ALICE} Collaboration, B.~Abelev {\em et~al.}, ``{Measurement of
  electrons from semileptonic heavy-flavor hadron decays in $pp$ collisions at
  $\sqrt{s} = 2.76$ TeV},''
  \href{http://dx.doi.org/10.1103/PhysRevD.91.012001}{{\em Phys. Rev.}
  {\bfseries D91} (2015) 012001},
\href{http://arxiv.org/abs/1405.4117}{{\ttfamily arXiv:1405.4117 [nucl-ex]}}.

\bibitem{Abelev:2012gx}
{\bfseries ALICE} Collaboration, B.~Abelev {\em et~al.}, ``{Measurement of
  prompt $J/\psi$ and beauty hadron production cross sections at mid-rapidity
  in $pp$ collisions at $\sqrt{s} = 7$ TeV},''
  \href{http://dx.doi.org/10.1007/JHEP11(2012)065}{{\em JHEP} {\bfseries 11}
  (2012) 065},
\href{http://arxiv.org/abs/1205.5880}{{\ttfamily arXiv:1205.5880 [hep-ex]}}.

\bibitem{Aaij:2012jd}
{\bfseries LHCb} Collaboration, R.~Aaij {\em et~al.}, ``{Measurement of the
  $B^\pm$ production cross-section in $pp$ collisions at $\sqrt{s}=7$ TeV},''
  \href{http://dx.doi.org/10.1007/JHEP04(2012)093}{{\em JHEP} {\bfseries 04}
  (2012) 093},
\href{http://arxiv.org/abs/1202.4812}{{\ttfamily arXiv:1202.4812 [hep-ex]}}.

\bibitem{Chatrchyan:2011vh}
{\bfseries CMS} Collaboration, S.~Chatrchyan {\em et~al.}, ``{Measurement of
  the Strange $B$ Meson Production Cross Section with J/Psi $\phi$ Decays in
  $pp$ Collisions at $\sqrt{s}=7$ TeV},''
  \href{http://dx.doi.org/10.1103/PhysRevD.84.052008}{{\em Phys. Rev.}
  {\bfseries D84} (2011) 052008},
  \href{http://arxiv.org/abs/1106.4048}{{\ttfamily arXiv:1106.4048 [hep-ex]}}.
and references therein.

\bibitem{Aad:2011sp}
{\bfseries ATLAS} Collaboration, G.~Aad {\em et~al.}, ``{Measurement of the
  differential cross-sections of inclusive, prompt and non-prompt $J/\psi$
  production in proton-proton collisions at $\sqrt{s}=7$ TeV},''
  \href{http://dx.doi.org/10.1016/j.nuclphysb.2011.05.015}{{\em Nucl. Phys.}
  {\bfseries B850} (2011) 387--444},
\href{http://arxiv.org/abs/1104.3038}{{\ttfamily arXiv:1104.3038 [hep-ex]}}.

\bibitem{Cacciari:1998it}
M.~Cacciari, M.~Greco, and P.~Nason, ``{The $p_T$ spectrum in heavy flavor
  hadroproduction},''
  \href{http://dx.doi.org/10.1088/1126-6708/1998/05/007}{{\em JHEP} {\bfseries
  05} (1998) 007},
\href{http://arxiv.org/abs/hep-ph/9803400}{{\ttfamily arXiv:hep-ph/9803400
  [hep-ph]}}.

\bibitem{Cacciari:2001td}
M.~Cacciari, S.~Frixione, and P.~Nason, ``{The $p_T$ spectrum in heavy flavor
  photoproduction},''
  \href{http://dx.doi.org/10.1088/1126-6708/2001/03/006}{{\em JHEP} {\bfseries
  03} (2001) 006},
\href{http://arxiv.org/abs/hep-ph/0102134}{{\ttfamily arXiv:hep-ph/0102134
  [hep-ph]}}.

\bibitem{Cacciari:2012ny}
M.~Cacciari, S.~Frixione, N.~Houdeau, M.~L. Mangano, P.~Nason, and G.~Ridolfi,
  ``{Theoretical predictions for charm and bottom production at the LHC},''
  \href{http://dx.doi.org/10.1007/JHEP10(2012)137}{{\em JHEP} {\bfseries 10}
  (2012) 137},
\href{http://arxiv.org/abs/1205.6344}{{\ttfamily arXiv:1205.6344 [hep-ph]}}.

\bibitem{Adare:2006nq}
{\bfseries PHENIX} Collaboration, A.~Adare {\em et~al.}, ``{Energy Loss and
  Flow of Heavy Quarks in Au+Au Collisions at $\sqrt{s_{NN}}=200$ GeV},''
  \href{http://dx.doi.org/10.1103/PhysRevLett.98.172301}{{\em Phys. Rev. Lett.}
  {\bfseries 98} (2007) 172301},
\href{http://arxiv.org/abs/nucl-ex/0611018}{{\ttfamily arXiv:nucl-ex/0611018
  [nucl-ex]}}.

\bibitem{Adare:2010de}
{\bfseries PHENIX} Collaboration, A.~Adare {\em et~al.}, ``{Heavy quark
  production in $p+p$ and energy loss and flow of heavy quarks in Au+Au
  collisions at $\sqrt{s_{NN}}=200$ GeV},''
  \href{http://dx.doi.org/10.1103/PhysRevC.84.044905}{{\em Phys. Rev.}
  {\bfseries C84} (2011) 044905},
\href{http://arxiv.org/abs/1005.1627}{{\ttfamily arXiv:1005.1627 [nucl-ex]}}.

\bibitem{Adam:2015nna}
{\bfseries ALICE} Collaboration, J.~Adam {\em et~al.}, ``{Centrality dependence
  of high-$p_{\rm T}$ D meson suppression in Pb--Pb collisions at $\sqrt{s_{\rm
  NN}}$ = 2.76 TeV},'' \href{http://dx.doi.org/10.1007/JHEP11(2015)205}{{\em
  JHEP} {\bfseries 11} (2015) 205},
\href{http://arxiv.org/abs/1506.06604}{{\ttfamily arXiv:1506.06604 [nucl-ex]}}.

\bibitem{Adam:2015sza}
{\bfseries ALICE} Collaboration, J.~Adam {\em et~al.}, ``{Transverse momentum
  dependence of D-meson production in Pb--Pb collisions at $
  \sqrt{{\mathrm{s}}_{\mathrm{NN}}}=$ 2.76 TeV},''
  \href{http://dx.doi.org/10.1007/JHEP03(2016)081}{{\em JHEP} {\bfseries 03}
  (2016) 081},
\href{http://arxiv.org/abs/1509.06888}{{\ttfamily arXiv:1509.06888 [nucl-ex]}}.

\bibitem{Abelev:2012qh}
{\bfseries ALICE} Collaboration, B.~Abelev {\em et~al.}, ``{Production of Muons
  from Heavy Flavor Decays at Forward Rapidity in pp and Pb--Pb Collisions at
  $\sqrt {s_{NN}}$ = 2.76 TeV},''
  \href{http://dx.doi.org/10.1103/PhysRevLett.109.112301}{{\em Phys. Rev.
  Lett.} {\bfseries 109} (2012) 112301},
\href{http://arxiv.org/abs/1205.6443}{{\ttfamily arXiv:1205.6443 [hep-ex]}}.

\bibitem{Abelev:2014laa}
{\bfseries ALICE} Collaboration, B.~Abelev {\em et~al.}, ``{Production of
  charged pions, kaons and protons at large transverse momenta in pp and
  Pb–Pb collisions at $\sqrt{s_{\rm NN}}$ =2.76 TeV},''
  \href{http://dx.doi.org/10.1016/j.physletb.2014.07.011}{{\em Phys. Lett.}
  {\bfseries B736} (2014) 196--207},
\href{http://arxiv.org/abs/1401.1250}{{\ttfamily arXiv:1401.1250 [nucl-ex]}}.

\bibitem{Djordjevic:2014tka}
M.~Djordjevic, M.~Djordjevic, and B.~Blagojevic, ``{RHIC and LHC jet
  suppression in non-central collisions},''
  \href{http://dx.doi.org/10.1016/j.physletb.2014.08.063}{{\em Phys. Lett.}
  {\bfseries B737} (2014) 298--302},
\href{http://arxiv.org/abs/1405.4250}{{\ttfamily arXiv:1405.4250 [nucl-th]}}.

\bibitem{Djordjevic:2013pba}
M.~Djordjevic, ``{Heavy flavor puzzle at LHC: a serendipitous interplay of jet
  suppression and fragmentation},''
  \href{http://dx.doi.org/10.1103/PhysRevLett.112.042302}{{\em Phys. Rev.
  Lett.} {\bfseries 112} (2014) 042302},
\href{http://arxiv.org/abs/1307.4702}{{\ttfamily arXiv:1307.4702 [nucl-th]}}.

\bibitem{Adam:2015rba}
{\bfseries ALICE} Collaboration, J.~Adam {\em et~al.}, ``{Inclusive, prompt and
  non-prompt J/$\psi$ production at mid-rapidity in Pb--Pb collisions at
  $\sqrt{s_{\rm NN}}$ = 2.76 TeV},''
  \href{http://dx.doi.org/10.1007/JHEP07(2015)051}{{\em JHEP} {\bfseries 07}
  (2015) 051},
\href{http://arxiv.org/abs/1504.07151}{{\ttfamily arXiv:1504.07151 [nucl-ex]}}.

\bibitem{Chatrchyan:2012np}
{\bfseries CMS} Collaboration, S.~Chatrchyan {\em et~al.}, ``{Suppression of
  non-prompt $J/\psi$, prompt $J/\psi$, and Y(1S) in PbPb collisions at
  $\sqrt{s_{NN}}=2.76$ TeV},''
  \href{http://dx.doi.org/10.1007/JHEP05(2012)063}{{\em JHEP} {\bfseries 05}
  (2012) 063},
\href{http://arxiv.org/abs/1201.5069}{{\ttfamily arXiv:1201.5069 [nucl-ex]}}.

\bibitem{Chatrchyan:2013exa}
{\bfseries CMS} Collaboration, S.~Chatrchyan {\em et~al.}, ``{Evidence of b-Jet
  Quenching in PbPb Collisions at $\sqrt{s_{NN}}=2.76$ TeV},''
  \href{http://dx.doi.org/10.1103/PhysRevLett.115.029903,
  10.1103/PhysRevLett.113.132301}{{\em Phys. Rev. Lett.} {\bfseries 113} (2014)
  132301}, \href{http://arxiv.org/abs/1312.4198}{{\ttfamily arXiv:1312.4198
  [nucl-ex]}}.
[Erratum: Phys. Rev. Lett.115,029903(2015)].

\bibitem{Adare:2015hla}
{\bfseries PHENIX} Collaboration, A.~Adare {\em et~al.}, ``{Single electron
  yields from semileptonic charm and bottom hadron decays in Au$+$Au collisions
  at $\sqrt{s_{NN}}=200$ GeV},''
  \href{http://dx.doi.org/10.1103/PhysRevC.93.034904}{{\em Phys. Rev.}
  {\bfseries C93} (2016) 034904},
\href{http://arxiv.org/abs/1509.04662}{{\ttfamily arXiv:1509.04662 [nucl-ex]}}.

\bibitem{Khachatryan:2015uja}
{\bfseries CMS} Collaboration, V.~Khachatryan {\em et~al.}, ``{Study of $B$
  Meson Production in $p+\mathrm{Pb}$ Collisions at $\sqrt{{s}_{NN}}=5.02\text{
  }\text{ }\mathrm{TeV}$ Using Exclusive Hadronic Decays},''
  \href{http://dx.doi.org/10.1103/PhysRevLett.116.032301}{{\em Phys. Rev.
  Lett.} {\bfseries 116} (2016) 032301},
\href{http://arxiv.org/abs/1508.06678}{{\ttfamily arXiv:1508.06678 [nucl-ex]}}.

\bibitem{CMS:2014tca}
{\bfseries CMS} Collaboration, V.~Khachatryan {\em et~al.}, ``{Transverse
  momentum spectra of inclusive b jets in pPb collisions at $\sqrt{s_{NN}} = $
  5.02 TeV},'' \href{http://dx.doi.org/10.1016/j.physletb.2016.01.010}{{\em
  Phys. Lett.} {\bfseries B754} (2016) 59},
\href{http://arxiv.org/abs/1510.03373}{{\ttfamily arXiv:1510.03373 [nucl-ex]}}.

\bibitem{CMS:2015qud}
{\bfseries CMS} Collaboration, ``{$J/\psi$ production in p Pb collisions},''.
CMS-PAS-HIN-14-009.

\bibitem{Aaij:2013zxa}
{\bfseries LHCb} Collaboration, R.~Aaij {\em et~al.}, ``{Study of $J/\psi$
  production and cold nuclear matter effects in $pPb$ collisions at
  $\sqrt{s_{NN}} = 5$ TeV},''
  \href{http://dx.doi.org/10.1007/JHEP02(2014)072}{{\em JHEP} {\bfseries 02}
  (2014) 072},
\href{http://arxiv.org/abs/1308.6729}{{\ttfamily arXiv:1308.6729 [nucl-ex]}}.

\bibitem{Abelev:2014hha}
{\bfseries ALICE} Collaboration, B.~Abelev {\em et~al.}, ``{Measurement of
  Prompt $D$-Meson Production in $p--Pb$ Collisions at $\sqrt{s_{NN}}$ = 5.02
  TeV},'' \href{http://dx.doi.org/10.1103/PhysRevLett.113.232301}{{\em Phys.
  Rev. Lett.} {\bfseries 113} (2014) 232301},
\href{http://arxiv.org/abs/1405.3452}{{\ttfamily arXiv:1405.3452 [nucl-ex]}}.

\bibitem{Adam:2015qda}
{\bfseries ALICE} Collaboration, J.~Adam {\em et~al.}, ``{Measurement of
  electrons from heavy-flavour hadron decays in p--Pb collisions at
  $\sqrt{s_{\rm NN}} = 5.02$ TeV},''
  \href{http://dx.doi.org/10.1016/j.physletb.2015.12.067}{{\em Phys. Lett.}
  {\bfseries B754} (2016) 81--93},
\href{http://arxiv.org/abs/1509.07491}{{\ttfamily arXiv:1509.07491 [nucl-ex]}}.

\bibitem{Adare:2013lkk}
{\bfseries PHENIX} Collaboration, A.~Adare {\em et~al.}, ``{Cold-Nuclear-Matter
  Effects on Heavy-Quark Production at Forward and Backward Rapidity in d+Au
  Collisions at $\sqrt{s_{NN}}$=200 GeV},''
  \href{http://dx.doi.org/10.1103/PhysRevLett.112.252301}{{\em Phys. Rev.
  Lett.} {\bfseries 112} (2014) 252301},
\href{http://arxiv.org/abs/1310.1005}{{\ttfamily arXiv:1310.1005 [nucl-ex]}}.

\bibitem{Abelev:2012ola}
{\bfseries ALICE} Collaboration, B.~Abelev {\em et~al.}, ``{Long-range angular
  correlations on the near and away side in $p$-Pb collisions at
  $\sqrt{s_{NN}}=5.02$ TeV},''
  \href{http://dx.doi.org/10.1016/j.physletb.2013.01.012}{{\em Phys. Lett.}
  {\bfseries B719} (2013) 29--41},
\href{http://arxiv.org/abs/1212.2001}{{\ttfamily arXiv:1212.2001 [nucl-ex]}}.

\bibitem{ABELEV:2013wsa}
{\bfseries ALICE} Collaboration, B.~Abelev {\em et~al.}, ``{Long-range angular
  correlations of $\rm \pi$, K and p in p--Pb collisions at $\sqrt{s_{\rm NN}}$
  = 5.02 TeV},'' \href{http://dx.doi.org/10.1016/j.physletb.2013.08.024}{{\em
  Phys. Lett.} {\bfseries B726} (2013) 164--177},
\href{http://arxiv.org/abs/1307.3237}{{\ttfamily arXiv:1307.3237 [nucl-ex]}}.

\bibitem{Aad:2013fja}
{\bfseries ATLAS} Collaboration, G.~Aad {\em et~al.}, ``{Measurement with the
  ATLAS detector of multi-particle azimuthal correlations in p+Pb collisions at
  $\sqrt{s_{\rm NN}}$=5.02 TeV},''
  \href{http://dx.doi.org/10.1016/j.physletb.2013.06.057}{{\em Phys. Lett.}
  {\bfseries B725} (2013) 60--78},
\href{http://arxiv.org/abs/1303.2084}{{\ttfamily arXiv:1303.2084 [hep-ex]}}.

\bibitem{CMS:2012qk}
{\bfseries CMS} Collaboration, S.~Chatrchyan {\em et~al.}, ``{Observation of
  long-range near-side angular correlations in proton-lead collisions at the
  LHC},'' \href{http://dx.doi.org/10.1016/j.physletb.2012.11.025}{{\em Phys.
  Lett.} {\bfseries B718} (2013) 795--814},
\href{http://arxiv.org/abs/1210.5482}{{\ttfamily arXiv:1210.5482 [nucl-ex]}}.

\bibitem{Adare:2013piz}
{\bfseries PHENIX} Collaboration, A.~Adare {\em et~al.}, ``{Quadrupole
  Anisotropy in Dihadron Azimuthal Correlations in Central $d$$+$Au Collisions
  at $\sqrt{s_{_{NN}}}$=200 GeV},''
  \href{http://dx.doi.org/10.1103/PhysRevLett.111.212301}{{\em Phys. Rev.
  Lett.} {\bfseries 111} (2013) 212301},
\href{http://arxiv.org/abs/1303.1794}{{\ttfamily arXiv:1303.1794 [nucl-ex]}}.

\bibitem{Adamczyk:2015xjc}
{\bfseries STAR} Collaboration, L.~Adamczyk {\em et~al.}, ``{Long-range
  pseudorapidity dihadron correlations in $d$+Au collisions at $\sqrt{s_{\rm
  NN}}=200$ GeV},''
  \href{http://dx.doi.org/10.1016/j.physletb.2015.05.075}{{\em Phys. Lett.}
  {\bfseries B747} (2015) 265--271},
\href{http://arxiv.org/abs/1502.07652}{{\ttfamily arXiv:1502.07652 [nucl-ex]}}.

\bibitem{Sickles:2013yna}
A.~M. Sickles, ``{Possible evidence for radial flow of heavy mesons in d+Au
  collisions},'' \href{http://dx.doi.org/10.1016/j.physletb.2014.02.013}{{\em
  Phys. Lett.} {\bfseries B731} (2014) 51--56},
\href{http://arxiv.org/abs/1309.6924}{{\ttfamily arXiv:1309.6924 [nucl-th]}}.

\bibitem{Dainese:2003zu}
A.~Dainese, {\em {Charm production and in-medium QCD energy loss in nucleus
  nucleus collisions with ALICE: A Performance study}}.
\newblock PhD thesis, Padua U., 2003.
\newblock
\href{http://arxiv.org/abs/nucl-ex/0311004}{{\ttfamily arXiv:nucl-ex/0311004
  [nucl-ex]}}.
\newblock

\bibitem{Aamodt:2008zz}
{\bfseries ALICE} Collaboration, K.~Aamodt {\em et~al.}, ``{The ALICE
  experiment at the CERN LHC},''
\href{http://dx.doi.org/10.1088/1748-0221/3/08/S08002}{{\em JINST} {\bfseries
  3} (2008) S08002}.

\bibitem{Abelev:2014ffa}
{\bfseries ALICE} Collaboration, B.~Abelev {\em et~al.}, ``{Performance of the
  ALICE experiment at the CERN LHC},''
  \href{http://dx.doi.org/10.1142/S0217751X14300440}{{\em Int. J. Mod. Phys.}
  {\bfseries A29} (2014) 1430044},
\href{http://arxiv.org/abs/1402.4476}{{\ttfamily arXiv:1402.4476 [nucl-ex]}}.

\bibitem{Aamodt:2010aa}
{\bfseries ALICE} Collaboration, K.~Aamodt {\em et~al.}, ``{Alignment of the
  ALICE Inner Tracking System with cosmic-ray tracks},''
  \href{http://dx.doi.org/10.1088/1748-0221/5/03/P03003}{{\em JINST} {\bfseries
  5} (2010) P03003},
\href{http://arxiv.org/abs/1001.0502}{{\ttfamily arXiv:1001.0502
  [physics.ins-det]}}.

\bibitem{Alme:2010ke}
J.~Alme {\em et~al.}, ``{The ALICE TPC, a large 3-dimensional tracking device
  with fast readout for ultra-high multiplicity events},''
  \href{http://dx.doi.org/10.1016/j.nima.2010.04.042}{{\em Nucl. Instrum.
  Meth.} {\bfseries A622} (2010) 316--367},
\href{http://arxiv.org/abs/1001.1950}{{\ttfamily arXiv:1001.1950
  [physics.ins-det]}}.

\bibitem{Cortese:519145}
{\bfseries ALICE} Collaboration, P.~Cortese {\em et~al.}, ``{ALICE
  Transition-Radiation Detector : Technical Design Report},''.
  \url{https://cds.cern.ch/record/519145}.
CERN-LHCC-2001-021.

\bibitem{Cortese:2002kf}
{\bfseries ALICE} Collaboration, P.~Cortese {\em et~al.}, ``{ALICE Time-Of
  Flight system (TOF) : addendum to the Technical Design Report},''.
  \url{https://cds.cern.ch/record/545834}.
CERN-LHCC-2002-016.

\bibitem{Cortese:2004aa}
{\bfseries ALICE} Collaboration, P.~Cortese {\em et~al.}, ``{ALICE Forward
  Detectors: FMD, TO and VO : Technical Design Report},''.
  \url{https://cds.cern.ch/record/781854}.
CERN-LHCC-2004-025.

\bibitem{Abelev:2013qoq}
{\bfseries ALICE} Collaboration, B.~Abelev {\em et~al.}, ``{Centrality
  determination of Pb--Pb collisions at $\sqrt{s_{NN}}$ = 2.76 TeV with
  ALICE},'' \href{http://dx.doi.org/10.1103/PhysRevC.88.044909}{{\em Phys.
  Rev.} {\bfseries C88} (2013) 044909},
\href{http://arxiv.org/abs/1301.4361}{{\ttfamily arXiv:1301.4361 [nucl-ex]}}.

\bibitem{Agashe:2014kda}
{\bfseries Particle Data Group} Collaboration, K.~A. Olive {\em et~al.},
  ``{Review of Particle Physics},''
\href{http://dx.doi.org/10.1088/1674-1137/38/9/090001}{{\em Chin. Phys.}
  {\bfseries C38} (2014) 090001}.

\bibitem{Gyulassy:1994ew}
M.~Gyulassy and X.-N. Wang, ``{HIJING 1.0: A Monte Carlo program for parton and
  particle production in high-energy hadronic and nuclear collisions},''
  \href{http://dx.doi.org/10.1016/0010-4655(94)90057-4}{{\em Comput. Phys.
  Commun.} {\bfseries 83} (1994) 307},
\href{http://arxiv.org/abs/nucl-th/9502021}{{\ttfamily arXiv:nucl-th/9502021
  [nucl-th]}}.

\bibitem{Sjostrand:2006za}
T.~Sjostrand, S.~Mrenna, and P.~Z. Skands, ``{PYTHIA 6.4 physics and manual},''
  \href{http://dx.doi.org/10.1088/1126-6708/2006/05/026}{{\em JHEP} {\bfseries
  05} (2006) 026},
\href{http://arxiv.org/abs/hep-ph/0603175}{{\ttfamily arXiv:hep-ph/0603175
  [hep-ph]}}.

\bibitem{Skands:2010ak}
P.~Z. Skands, ``{Tuning Monte Carlo generators: The Perugia tunes},''
  \href{http://dx.doi.org/10.1103/PhysRevD.82.074018}{{\em Phys. Rev.}
  {\bfseries D82} (2010) 074018},
\href{http://arxiv.org/abs/1005.3457}{{\ttfamily arXiv:1005.3457 [hep-ph]}}.

\bibitem{Brun:1994aa}
R.~Brun {\em et~al.}, ``{GEANT: Detector Description and Simulation Tool},''.
  \url{https://cds.cern.ch/record/1082634}.
CERN-W-5013.

\bibitem{Abelev:2012xe}
{\bfseries ALICE} Collaboration, B.~Abelev {\em et~al.}, ``{Measurement of
  electrons from semileptonic heavy-flavour hadron decays in pp collisions at
  $\sqrt{s}$ = 7 TeV},''
  \href{http://dx.doi.org/10.1103/PhysRevD.86.112007}{{\em Phys. Rev.}
  {\bfseries D86} (2012) 112007},
\href{http://arxiv.org/abs/1205.5423}{{\ttfamily arXiv:1205.5423 [hep-ex]}}.

\bibitem{Abelev:2013haa}
{\bfseries ALICE} Collaboration, B.~Abelev {\em et~al.}, ``{Multiplicity
  dependence of pion, kaon, proton and lambda production in p--Pb collisions at
  $\mathbf{\sqrt{{\textit s}_{\rm NN}}}$ = 5.02 TeV},''
  \href{http://dx.doi.org/10.1016/j.physletb.2013.11.020}{{\em Phys. Lett.}
  {\bfseries B728} (2014) 25--38},
\href{http://arxiv.org/abs/1307.6796}{{\ttfamily arXiv:1307.6796 [nucl-ex]}}.

\bibitem{Adam:2015pionshighpt}
{\bfseries ALICE} Collaboration, J.~Adam {\em et~al.}, ``{Multiplicity
  dependence of charged pion, kaon, and (anti)proton production at large
  transverse momentum in p--Pb collisions at $\mathbf{\sqrt{{\textit s}_{\rm
  NN}}}$ = 5.02 TeV},''
\href{http://arxiv.org/abs/1601.03658}{{\ttfamily arXiv:1601.03658 [nucl-ex]}}.

\bibitem{Abramowicz:2013eja}
{\bfseries ZEUS} Collaboration, H.~Abramowicz {\em et~al.}, ``{Measurement of
  charm fragmentation fractions in photoproduction at HERA},''
  \href{http://dx.doi.org/10.1007/JHEP09(2013)058}{{\em JHEP} {\bfseries 09}
  (2013) 058},
\href{http://arxiv.org/abs/1306.4862}{{\ttfamily arXiv:1306.4862 [hep-ex]}}.

\bibitem{CowanStatistical}
{G. Cowan}, {\em {Statistical data analysis}}.
\newblock {Oxford science publications}. {Clarendon Press}, {Oxford},
  {Repr.}~ed., {2004}.

\bibitem{ALICE:2012xs}
{\bfseries ALICE} Collaboration, B.~Abelev {\em et~al.}, ``{Pseudorapidity
  Density of Charged Particles in p--Pb Collisions at $\mathbf{\sqrt{{\textit
  s}_{\rm NN}}}$=5.02 TeV},''
  \href{http://dx.doi.org/10.1103/PhysRevLett.110.032301}{{\em Phys. Rev.
  Lett.} {\bfseries 110} (2013) 032301},
\href{http://arxiv.org/abs/1210.3615}{{\ttfamily arXiv:1210.3615 [nucl-ex]}}.

\bibitem{Abelev:2014epa}
{\bfseries ALICE} Collaboration, B.~Abelev {\em et~al.}, ``{Measurement of
  visible cross sections in proton-lead collisions at $\sqrt{s_{\rm NN}}$ =
  5.02 TeV in van der Meer scans with the ALICE detector},''
  \href{http://dx.doi.org/10.1088/1748-0221/9/11/P11003}{{\em JINST} {\bfseries
  9} (2014) P11003},
\href{http://arxiv.org/abs/1405.1849}{{\ttfamily arXiv:1405.1849 [nucl-ex]}}.

\bibitem{Abelev:2014dsa}
{\bfseries ALICE} Collaboration, B.~Abelev {\em et~al.}, ``{Transverse momentum
  dependence of inclusive primary charged-particle production in p--Pb
  collisions at $\sqrt{s_\mathrm{{NN}}}=5.02~\text {TeV}$},''
  \href{http://dx.doi.org/10.1140/epjc/s10052-014-3054-5}{{\em Eur. Phys. J.}
  {\bfseries C74} (2014) 3054},
\href{http://arxiv.org/abs/1405.2737}{{\ttfamily arXiv:1405.2737 [nucl-ex]}}.

\bibitem{PhysRevLett.111.222301}
{\bfseries ALICE} Collaboration, B.~Abelev {\em et~al.}, ``{$K^0_S$ and
  $\Lambda$ Production in Pb--Pb Collisions at $\sqrt{s_{NN}}$ = 2.76 TeV},''
  \href{http://dx.doi.org/10.1103/PhysRevLett.111.222301}{{\em Phys. Rev.
  Lett.} {\bfseries 111} (2013) 222301},
\href{http://arxiv.org/abs/1307.5530}{{\ttfamily arXiv:1307.5530 [nucl-ex]}}.

\bibitem{Barlow1993219}
R.~Barlow and C.~Beeston, ``{Fitting using finite Monte Carlo samples},'' {\em
  Comp. Phys. Comm.} {\bfseries 77} no.~2, (1993) 219.

\bibitem{Abelev:2014ypa}
{\bfseries ALICE} Collaboration, B.~Abelev {\em et~al.}, ``{Neutral pion
  production at midrapidity in pp and Pb--Pb collisions at $\sqrt{s_{{\mathrm
  {NN}}}}= 2.76\,{\mathrm {TeV}}$},''
  \href{http://dx.doi.org/10.1140/epjc/s10052-014-3108-8}{{\em Eur. Phys. J.}
  {\bfseries C74} (2014) 3108},
\href{http://arxiv.org/abs/1405.3794}{{\ttfamily arXiv:1405.3794 [nucl-ex]}}.

\bibitem{ALICE:2012ab}
{\bfseries ALICE} Collaboration, B.~Abelev {\em et~al.}, ``{Suppression of high
  transverse momentum D mesons in central Pb--Pb collisions at
  $\sqrt{s_{NN}}=2.76$ TeV},''
  \href{http://dx.doi.org/10.1007/JHEP09(2012)112}{{\em JHEP} {\bfseries 09}
  (2012) 112},
\href{http://arxiv.org/abs/1203.2160}{{\ttfamily arXiv:1203.2160 [nucl-ex]}}.

\bibitem{He:2014cla}
M.~He, R.~J. Fries, and R.~Rapp, ``{Heavy flavor at the large hadron collider
  in a strong coupling approach},''
  \href{http://dx.doi.org/10.1016/j.physletb.2014.05.050}{{\em Phys. Lett.}
  {\bfseries B735} (2014) 445--450},
\href{http://arxiv.org/abs/1401.3817}{{\ttfamily arXiv:1401.3817 [nucl-th]}}.

\bibitem{Oh:2009zj}
Y.~Oh, C.~M. Ko, S.~H. Lee, and S.~Yasui, ``{Heavy baryon/meson ratios in
  relativistic heavy ion collisions},''
  \href{http://dx.doi.org/10.1103/PhysRevC.79.044905}{{\em Phys. Rev.}
  {\bfseries C79} (2009) 044905},
\href{http://arxiv.org/abs/0901.1382}{{\ttfamily arXiv:0901.1382 [nucl-th]}}.

\bibitem{Averbeck:2011ga}
R.~Averbeck, N.~Bastid, Z.~C. del Valle, P.~Crochet, A.~Dainese, and X.~Zhang,
  ``{Reference heavy flavour cross sections in pp collisions at $\sqrt{s}$ =
  2.76 TeV, using a pQCD-driven $\sqrt{s}$-scaling of ALICE measurements at
  $\sqrt{s}$ = 7 TeV},''
\href{http://arxiv.org/abs/1107.3243}{{\ttfamily arXiv:1107.3243 [hep-ph]}}.

\bibitem{Nadolsky:2008zw}
P.~M. Nadolsky, H.-L. Lai, Q.-H. Cao, J.~Huston, J.~Pumplin, D.~Stump, W.-K.
  Tung, and C.~P. Yuan, ``{Implications of CTEQ global analysis for collider
  observables},'' \href{http://dx.doi.org/10.1103/PhysRevD.78.013004}{{\em
  Phys. Rev.} {\bfseries D78} (2008) 013004},
\href{http://arxiv.org/abs/0802.0007}{{\ttfamily arXiv:0802.0007 [hep-ph]}}.

\bibitem{Abelev:2014hla}
{\bfseries ALICE} Collaboration, B.~Abelev {\em et~al.}, ``{Beauty production
  in pp collisions at $\sqrt{s}$ = 2.76 TeV measured via semi-electronic
  decays},'' \href{http://dx.doi.org/10.1016/j.physletb.2016.10.004,
  10.1016/j.physletb.2014.09.026}{{\em Phys. Lett.} {\bfseries B738} (2014)
  97--108}, \href{http://arxiv.org/abs/1405.4144}{{\ttfamily arXiv:1405.4144
  [nucl-ex]}}.
[Erratum: Phys. Lett.B763,507(2016)].

\bibitem{InclusiveelectronRAA}
{\bfseries ALICE} Collaboration, J.~Adam {\em et~al.}, ``{Measurement of the
  production of high-$p_{\rm T}$ electrons from heavy-flavour hadron decays in
  Pb-Pb collisions at $\mathbf{\sqrt{\it s_{\rm{NN}}}}$ = 2.76 TeV},''
\href{http://arxiv.org/abs/1609.07104}{{\ttfamily arXiv:1609.07104 [nucl-ex]}}.

\bibitem{Eskola:2009uj}
K.~J. Eskola, H.~Paukkunen, and C.~A. Salgado, ``{EPS09 -- A new generation of
  NLO and LO nuclear parton distribution functions},''
  \href{http://dx.doi.org/10.1088/1126-6708/2009/04/065}{{\em JHEP} {\bfseries
  04} (2009) 065},
\href{http://arxiv.org/abs/0902.4154}{{\ttfamily arXiv:0902.4154 [hep-ph]}}.

\bibitem{Nahrgang:2013xaa}
M.~Nahrgang, J.~Aichelin, P.~B. Gossiaux, and K.~Werner, ``{Influence of
  hadronic bound states above $T_c$ on heavy-quark observables in Pb + Pb
  collisions at at the CERN Large Hadron Collider},''
  \href{http://dx.doi.org/10.1103/PhysRevC.89.014905}{{\em Phys. Rev.}
  {\bfseries C89} (2014) 014905},
\href{http://arxiv.org/abs/1305.6544}{{\ttfamily arXiv:1305.6544 [hep-ph]}}.

\bibitem{Beraudo:2014boa}
A.~Beraudo, A.~De~Pace, M.~Monteno, M.~Nardi, and F.~Prino, ``{Heavy flavors in
  heavy-ion collisions: quenching, flow and correlations},''
  \href{http://dx.doi.org/10.1140/epjc/s10052-015-3336-6}{{\em Eur. Phys. J.}
  {\bfseries C75} (2015) 121},
\href{http://arxiv.org/abs/1410.6082}{{\ttfamily arXiv:1410.6082 [hep-ph]}}.

\bibitem{Horowitz:2011gd}
W.~A. Horowitz and M.~Gyulassy, ``{The surprisingly transparent sQGP at LHC},''
  \href{http://dx.doi.org/10.1016/j.nuclphysa.2011.09.018}{{\em Nucl. Phys.}
  {\bfseries A872} (2011) 265--285},
\href{http://arxiv.org/abs/1104.4958}{{\ttfamily arXiv:1104.4958 [hep-ph]}}.

\bibitem{Wicks:2005gt}
S.~Wicks, W.~Horowitz, M.~Djordjevic, and M.~Gyulassy, ``{Elastic, inelastic,
  and path length fluctuations in jet tomography},''
  \href{http://dx.doi.org/10.1016/j.nuclphysa.2006.12.048}{{\em Nucl. Phys.}
  {\bfseries A784} (2007) 426--442},
\href{http://arxiv.org/abs/nucl-th/0512076}{{\ttfamily arXiv:nucl-th/0512076
  [nucl-th]}}.

\bibitem{Horowitz:2015dta}
W.~A. Horowitz, ``{Fluctuating heavy quark energy loss in a strongly coupled
  quark-gluon plasma},''
  \href{http://dx.doi.org/10.1103/PhysRevD.91.085019}{{\em Phys. Rev.}
  {\bfseries D91} (2015) 085019},
\href{http://arxiv.org/abs/1501.04693}{{\ttfamily arXiv:1501.04693 [hep-ph]}}.

\bibitem{Uphoff:2014hza}
J.~Uphoff, O.~Fochler, Z.~Xu, and C.~Greiner, ``{Elastic and radiative heavy
  quark interactions in ultra-relativistic heavy-ion collisions},''
  \href{http://dx.doi.org/10.1088/0954-3899/42/11/115106}{{\em J. Phys.}
  {\bfseries G42} (2015) 115106},
\href{http://arxiv.org/abs/1408.2964}{{\ttfamily arXiv:1408.2964 [hep-ph]}}.

\bibitem{Uphoff:2014cba}
J.~Uphoff, F.~Senzel, O.~Fochler, C.~Wesp, Z.~Xu, and C.~Greiner, ``{Elliptic
  Flow and Nuclear Modification Factor in Ultrarelativistic Heavy-Ion
  Collisions within a Partonic Transport Model},''
  \href{http://dx.doi.org/10.1103/PhysRevLett.114.112301}{{\em Phys. Rev.
  Lett.} {\bfseries 114} (2015) 112301},
\href{http://arxiv.org/abs/1401.1364}{{\ttfamily arXiv:1401.1364 [hep-ph]}}.

\bibitem{Werner:2010aa}
K.~Werner, I.~Karpenko, T.~Pierog, M.~Bleicher, and K.~Mikhailov,
  ``{Event-by-event simulation of the three-dimensional hydrodynamic evolution
  from flux tube initial conditions in ultrarelativistic heavy ion
  collisions},'' \href{http://dx.doi.org/10.1103/PhysRevC.82.044904}{{\em Phys.
  Rev.} {\bfseries C82} (2010) 044904},
\href{http://arxiv.org/abs/1004.0805}{{\ttfamily arXiv:1004.0805 [nucl-th]}}.

\bibitem{Werner:2012xh}
K.~Werner, I.~Karpenko, M.~Bleicher, T.~Pierog, and S.~Porteboeuf-Houssais,
  ``{Jets, bulk matter, and their interaction in heavy ion collisions at
  several TeV},'' \href{http://dx.doi.org/10.1103/PhysRevC.85.064907}{{\em
  Phys. Rev.} {\bfseries C85} (2012) 064907},
\href{http://arxiv.org/abs/1203.5704}{{\ttfamily arXiv:1203.5704 [nucl-th]}}.

\bibitem{Baier:1994bd}
R.~Baier, Y.~L. Dokshitzer, S.~Peigne, and D.~Schiff, ``{Induced gluon
  radiation in a QCD medium},''
  \href{http://dx.doi.org/10.1016/0370-2693(94)01617-L}{{\em Phys. Lett.}
  {\bfseries B345} (1995) 277--286},
\href{http://arxiv.org/abs/hep-ph/9411409}{{\ttfamily arXiv:hep-ph/9411409
  [hep-ph]}}.

\bibitem{Braaten:1989mz}
E.~Braaten and R.~D. Pisarski, ``{Soft amplitudes in hot gauge theories: A
  general analysis},''
\href{http://dx.doi.org/10.1016/0550-3213(90)90508-B}{{\em Nucl. Phys.}
  {\bfseries B337} (1990) 569--634}.

\end{thebibliography}\endgroup

\newpage
\appendix
\section{The ALICE Collaboration}
\label{app:collab}

\bigskip 

J.~Adam$^{\rm 39,88}$, 
D.~Adamov\'{a}$^{\rm 85}$, 
M.M.~Aggarwal$^{\rm 89}$, 
G.~Aglieri Rinella$^{\rm 35}$, 
M.~Agnello$^{\rm 31,112}$, 
N.~Agrawal$^{\rm 48}$, 
Z.~Ahammed$^{\rm 136}$, 
S.~Ahmad$^{\rm 18}$, 
S.U.~Ahn$^{\rm 69}$, 
S.~Aiola$^{\rm 140}$
% ...

\bigskip

\bigskip 

\textbf{\Large Affiliation Notes}

\bigskip 

% ...
$^{\rm IV}$ Also at: M.V. Lomonosov Moscow State University, D.V. Skobeltsyn Institute of Nuclear, Physics, Moscow, Russia\\

\bigskip

\bigskip 

\textbf{\Large Collaboration Institutes}

\bigskip 

% ...
$^{18}$ Department of Physics, Aligarh Muslim University, Aligarh, India\\
% ...
$^{31}$ Dipartimento DISAT del Politecnico and Sezione INFN, Turin, Italy\\
% ...
$^{35}$ European Organization for Nuclear Research (CERN), Geneva, Switzerland\\
% ...
$^{39}$ Faculty of Nuclear Sciences and Physical Engineering, Czech Technical University in Prague, Prague, Czech Republic\\
% ...
$^{48}$ Indian Institute of Technology Bombay (IIT), Mumbai, India\\
% ...
$^{69}$ Korea Institute of Science and Technology Information, Daejeon, South Korea\\
% ...
$^{85}$ Nuclear Physics Institute, Academy of Sciences of the Czech Republic, \v{R}e\v{z} u Prahy, Czech Republic\\
% ...
$^{88}$ Physics Department, Creighton University, Omaha, Nebraska, United States\\
$^{89}$ Physics Department, Panjab University, Chandigarh, India\\
% ...
$^{112}$ Sezione INFN, Turin, Italy\\
% ...
$^{136}$ Variable Energy Cyclotron Centre, Kolkata, India\\
% ...
$^{140}$ Yale University, New Haven, Connecticut, United States\\
% ...

%
%
\end{document}